\documentclass[prd,preprint,tightenlines,floatfix,showpacs,preprintnumbers,nofootinbib,eqsecnum]{revtex4}

 \usepackage[dvips,final]{graphicx}
  \usepackage{amssymb}
   \usepackage{amsmath}
    \usepackage{amsfonts}
     \usepackage{epsfig}
      \usepackage{bm}% bold math

\usepackage[section]{placeins}

\usepackage{multirow}
\usepackage{ctable}
\usepackage{booktabs}
\usepackage{array}
\usepackage{tabularx}
\usepackage{xcolor}
\usepackage{pstricks}
\begin{document}

\vfill
\title{Production of \bm{$c \bar c c\bar c$} in double-parton scattering 
within \bm{$k_{t}$}-factorization approach -- meson-meson correlations}

\author{Rafa{\l} Maciu{\l}a}
\email{rafal.maciula@ifj.edu.pl} \affiliation{Institute of Nuclear
Physics PAN, PL-31-342 Cracow, Poland}

\author{Antoni Szczurek}
\email{antoni.szczurek@ifj.edu.pl} \affiliation{Institute of Nuclear
Physics PAN, PL-31-342 Cracow,
Poland and\\
University of Rzesz\'ow, PL-35-959 Rzesz\'ow, Poland}

\date{\today}

\begin{abstract}
We discuss production of two pairs of $c \bar c$ in proton-proton collisions at the LHC.
Both double-parton scattering (DPS) and single-parton scattering (SPS)
contributions are included in the analysis. Each step of DPS is
calculated within $k_t$-factorization approach, i.e. effectively including next-to-leading order corrections. The conditions how
to identify the DPS contribution are presented. The discussed mechanism
unavoidably leads to the production of pairs of mesons:
$D_i D_j$ (each containing $c$ quarks) or $\bar D_i \bar D_j$ (each containing $\bar c$ antiquarks). We calculate corresponding production rates
for different combinations of charmed mesons as well as some differential distribution for
$(D^0 D^0$ + $\bar D^0 \bar D^0)$ production. Within large theoretical uncertainties the predicted DPS cross section is
fairly similar to the cross section measured recently by the LHCb collaboration.
The best description is obtained with the Kimber-Martin-Ryskin (KMR)
unintegrated gluon distribution, which very well simulates
higher-order corrections. The contribution of SPS,
calculated in the high-energy approximation, turned out to be rather
small. Finally, we emphasize significant contribution of DPS mechanism to inclusive charmed meson spectra measured recently by ALICE, ATLAS and LHCb.

\end{abstract}


\maketitle

%----------------------------
\section{Introduction}
%----------------------------

There has been recently renewed interest in studying double-parton scattering (DPS) effects in different reactions (see e.g. \cite{Bartalini2011} and references therein).
Very recently we have shown that the production of $c \bar c c \bar c$
is a very good place to study DPS effects \cite{LMS2012}.
Here, the quark mass is small enough to assure that the cross section for DPS is very large,
and large enough that each of the scatterings can be treated
within pQCD. The calculation performed in
Ref.~\cite{LMS2012} were done in the leading-order (LO) collinear approximation. This may not be 
sufficient when comparing the results of the calculation with real
experimental data. In the meantime the LHCb collaboration presented
new interesting data for simultaneous production of two charmed mesons \cite{LHCb-DPS-2012}. 
They have observed large percentage of the events with two mesons, 
both containing $c$ quark, with respect to the typical production of the corresponding 
meson/antimeson pair ($\sigma_{D_{i}D_{j}}/ \sigma_{D_{i}\bar{D_{j}}} \sim 10\%$), despite of the very limited LHCb acceptance. 

Is the large effect a footprint of double parton scattering? We wish to
address the issue in this paper. In addition, we shall
estimate $c \bar c c \bar c$ production via single-parton scattering (SPS) within a high-energy approximation \cite{SS2012}.
This approach should be an efficient tool especially when the distance in rapidity
between $c c$ or/and $\bar c \bar c$ is large.

Another evidence for the DPS effects can be a missing cross section
in the inclusive charmed meson distributions observed recently in Ref.~\cite{MS2013-charmed-meson}.
The measured inclusive cross sections include events where two $D$ (or two $\bar D$) mesons are produced, therefore corresponding theoretical predictions should also be corrected for the DPS effects.

In Ref.~\cite{Berezhnoy2012} the authors estimated DPS contribution
based on the experimental inclusive $D$ meson spectra measured at LHC 
which, as discussed in our paper, may be too crude approximation.
In addition, in their approach fragmentation was included only in terms
of the branching fractions for the transition $c \to D$. In our approach
we shall include full kinematics of hadronization process.
Here we wish to show also first differential distributions
on the hadron level to be confronted with recent LHCb experimental data \cite{LHCb-DPS-2012}.

%----------------------------------------
\section{Theoretical framework}
%----------------------------------------

\label{sec:theory}

In the present analysis, when considering $pp \to c \bar c c \bar c X$ reaction, we concentrate primarily on double-parton scattering
effects. In Section III.B we will show that the single-scattering contribution to double-charm production is much smaller, especially in the LHCb kinematics.

%-------------------------------------------------
\subsection{Double-parton scattering}
%-------------------------------------------------

In LO collinear approximation the differential distributions for $c\bar{c}$ production
depend e.g. on rapidity of quark, rapidity of antiquark and transverse
momentum of one of them (they are identical) \cite{LMS2012}. 
In the next-to-leading order (NLO) collinear approach or in the
$k_t$-factorization approach the situation is more complicated as there are more 
kinematical variables necessary to describe the kinematical situation.
In the $k_t$-factorization approach the differential cross section for
DPS production of $c \bar c c \bar c$ system, assuming factorization of the DPS model, can be written as: 
\begin{eqnarray}
\frac{d \sigma^{DPS}(p p \to c \bar c c \bar c X)}{d y_1 d y_2 d^2 p_{1,t} d^2 p_{2,t} 
d y_3 d y_4 d^2 p_{3,t} d^2 p_{4,t}} = \nonumber \;\;\;\;\;\;\;\;\;\;\;\;\;\;\;\;\;\;\;\;\;\;\;\;\;\;\;\;\;\;\;\;\;\;\;\;\;\;\;\;\;\;\;\;\;\;\;\;\;\;\;\;\;\;\;\;\;\;\;\;\;\;\;\;\;\;\; \\ 
\frac{1}{2 \sigma_{eff}} \cdot
\frac{d \sigma^{SPS}(p p \to c \bar c X_1)}{d y_1 d y_2 d^2 p_{1,t} d^2 p_{2,t}}
\cdot
\frac{d \sigma^{SPS}(p p \to c \bar c X_2)}{d y_3 d y_4 d^2 p_{3,t} d^2 p_{4,t}}.
\end{eqnarray}

When integrating over kinematical variables one obtains
\begin{equation}
\sigma^{DPS}(p p \to c \bar c c \bar c X) = \frac{1}{2 \sigma_{eff}}
\sigma^{SPS}(p p \to c \bar c X_1) \cdot \sigma^{SPS}(p p \to c \bar c X_2).
\label{basic_formula}
\end{equation}
These formulae assume that the two parton subprocesses are not correlated one
with each other. 
The parameter $\sigma_{eff}$ in the denominator of above formulae can be defined as:
\begin{equation}
\sigma_{eff} = \left[ \int d^{2}b (T(\vec{b}))^{2} \right]^{-1},
\end{equation}
where the overlap function
\begin{equation}
T ( \vec{b} ) = \int f( \vec{b}_{1} ) f(\vec{b}_{1} - \vec{b} ) d^2 b_{1},
\end{equation}
if the impact-parameter dependent double-parton distributions (dPDFs) are written
in the following factorized approximation \cite{GS2010,Gustaffson2011}:
\begin{equation}
\Gamma_{i,j}(x_1,x_2;\vec{b}_{1},\vec{b}_{2};\mu_{1}^{2},\mu_{2}^{2}) = F_{i,j}(x_1,x_2;\mu_{1}^{2},\mu_{2}^{2}) f(\vec{b}_{1}) f(\vec{b}_{2}).
\end{equation}
Experimental data from Tevatron \cite{Tevatron} provide an estimate of 
$\sigma_{eff}$ in the denominator of formula (\ref{basic_formula}). Corresponding evaluations from the LHC are expected soon. 
In our analysis we take $\sigma_{eff}$ = 15 mb. In the most general case one 
may expect some violation of this simple factorized Ansatz given by Eq.~\ref{basic_formula} \cite{Gustaffson2011}.

In our present analysis cross section for each step is calculated in the
$k_t$-factorization approach, that is:
\begin{eqnarray}
\frac{d \sigma^{SPS}(p p \to c \bar c X_1)}{d y_1 d y_2 d^2 p_{1,t} d^2 p_{2,t}} 
&& = \frac{1}{16 \pi^2 {\hat s}^2} \int \frac{d^2 k_{1t}}{\pi} \frac{d^2 k_{2t}}{\pi} \overline{|{\cal M}_{g^{*} g^{*} \rightarrow c \bar{c}}|^2} \nonumber \\
&& \times \;\; \delta^2 \left( \vec{k}_{1t} + \vec{k}_{2t} - \vec{p}_{1t} - \vec{p}_{2t}
\right)
{\cal F}(x_1,k_{1t}^2,\mu^2) {\cal F}(x_2,k_{2t}^2,\mu^2),
\nonumber
\end{eqnarray}
\begin{eqnarray}
\frac{d \sigma^{SPS}(p p \to c \bar c X_2)}{d y_3 d y_4 d^2 p_{3,t} d^2 p_{4,t}} 
&& = \frac{1}{16 \pi^2 {\hat s}^2} \int \frac{d^2 k_{3t}}{\pi} \frac{d^2 k_{4t}}{\pi} \overline{|{\cal M}_{g^{*} g^{*} \rightarrow c \bar{c}}|^2} \nonumber \\
&&\times \;\; \delta^2 \left( \vec{k}_{3t} + \vec{k}_{4t} - \vec{p}_{3t} - \vec{p}_{4t}
\right)
{\cal F}(x_3,k_{3t}^2,\mu^2) {\cal F}(x_4,k_{4t}^2,\mu^2).
\end{eqnarray}
The matrix elements for $g^* g^* \to c \bar c$ (off-shell gluons) must be calculated including transverse momenta of initial gluons as it was done first in \cite{CCH91,CE91,BE01}.
The unintegrated ($k_t$-dependent) gluon distributions (UGDFs) in the proton are taken from the literature \cite{KMR,KMS,Jung}. Due to the emision of soft gluons encoded in these objects, it is belived that a major part of NLO corrections is effectively included. This is in analogy to initial state parton shower in Monte Carlo generators and strongly depends on technical construction of UGDF (see Ref.~\cite{MS2013-charmed-meson}). 
The framework of the $k_t$-factorization approach is often used with success in describing inclusive spectra of
$D$ or $B$ mesons as well as for theoretical predictions for so-called nonphotonic leptons, products of semileptonic decays of charm and bottom mesons \cite{Teryaev,Baranov00,BLZ,Shuvaev,LMS09,MSS2011,JKLZ}.

%-------------------------------------------------
\subsection{Single-parton scattering}
%-------------------------------------------------

The total cross section for the production of $c \bar c c \bar c$ final state via single gluon-gluon interaction can be calculated in the parton model approach as:
\begin{equation}
\sigma(p p \to c \bar c c \bar c;W^2) = 
\int d x_1 d x_2 \; g(x_1,\mu_F^2) \; g(x_2,\mu_F^2) \;
 \sigma(gg \to c \bar c c \bar c;x_1 x_2 W^2) \; .
\label{parton_model}
\end{equation}
Here $g(x,\mu^2)$ is integrated (collinear) gluon distribution in a proton (PDF), and $W$ is the
proton-proton center of mass energy.
In practice the integration is done in $\log_{10}x_1$ and
$\log_{10}x_2$, including the corresponding jacobian of transformation.
The elementary cross section
of Eq. (\ref{parton_model}) enters at 
$\hat{s} = x_1 x_2 W^2 > 16 m_c^2$.
The parton level cross section in (\ref{parton_model}) is therefore very useful in order to 
obtain differential distributions in invariant mass of the $c \bar c c \bar c$ system. 

In the present calculation we concentrate on LHC energies and consider 
the gg $\to c \bar c c \bar c$ subprocesses only.
In the high-energy approximation the elementary cross section can be
written in the compact form (see Ref.~\cite{SS2012}):
\begin{eqnarray}
d \sigma(gg \to c \bar c c \bar c)=  
{N_c^2-1 \over N_c^2} \, 
{4\pi^2 \alpha_s^2 \over [\vec{q}^2 + \mu_G^2]^2} \,
I_{g \to c \bar c}(z_1,\vec{k}_1,\vec{q}) 
I_{g \to c \bar c}(z_2,\vec{k}_2,-\vec{q}) \, 
dz_1 {d^2{ k}_1 \over (2 \pi)^2} \, dz_2 {d^2 { k}_2 \over (2 \pi)^2} 
{d^2 {q} \over (2 \pi)^2}. \nonumber \\
\label{sigma_momentum_space}
\end{eqnarray}
Here the $I_{g \to c \bar c}(z_1,\vec{k}_1, \vec{ q}_1)$ and 
$I_{g \to c \bar c}(z_2,\vec{ k}_2, \vec{ q}_2)$ factors, called impact factors,
describe the coupling of pairs of $c \bar c$ associated with the first
and second gluon/proton, respectively.
Above $z_1$ and $z_2$ are longitudinal momentum fractions of quarks 
with respect to parent gluons in the first and second pair, respectively, 
and $\vec{k}_i$ their respective transverse momenta, $\vec{q}$ is exchanged transverse momentum and $\mu_G$ is gluon mass 
which can be put to zero at least mathematically.
At low energies this formula must be corrected for threshold effects \cite{SS2012}.
The differential cross sections for $p p \to c \bar c c \bar c X$ 
can be obtained by replacing
the $\sigma(g g \to c \bar c c \bar c)$ by 
  $d \sigma(g g \to c \bar c c \bar c)$ in Eq.(\ref{parton_model}).
Details about how the arguments of $\alpha_s$ are chosen are discussed in Ref.~\cite{SS2012}.

Our approach includes subprocesses coherently to be contrasted to Ref.~\cite{Berezhnoy2012}
where they were separated one from each other to simplify calculations. In addition, we get a practical agreement
with results of calculations in Ref.~\cite{Barger91}.

%-----------------------------------------------
\subsection{Double meson production}
%-----------------------------------------------

Kinematical correlations between quarks and antiquarks are not accessible
experimentally. Instead one can measure correlations between heavy mesons
or nonphotonic electrons. In this paper we will analyze kinematical correlations
between charmed mesons. In particular, we are interested in correlations
between $D_i$ and $D_j$ mesons (both containing $c$ quark) or between
$\bar D_i$ and $\bar D_j$ mesons (both containing $\bar c$ antiquark).
In order to calculate correlations between mesons we follow here the fragmentation function technique for hadronization process:
\begin{equation}
\frac{d \sigma(pp \to DDX)}{d y_1 d y_{2} d^2 p_{1,t}^{D} d^2 p_{2,t}^{D}}
 \approx
\int \frac{D_{c \to D}(z_{1})}{z_{1}}\cdot \frac{D_{c \to D}(z_{2})}{z_{2}}\cdot
\frac{d \sigma(pp \to ccX)}{d y_1 d y_{2} d^2
  p_{1,t}^{c} d^2 p_{2,t}^{c}} d z_{1} d z_{2} \; ,
\end{equation}
where: 
$p_{1,t}^{c} = \frac{p_{1,t}^{D}}{z_{1}}$, $p_{2,t}^{c} =
  \frac{p_{2,t}^{D}}{z_{2}}$ and
meson longitudinal fractions  $z_{1}, z_{2}\in (0,1)$.
We have made approximation assuming that $y_{1}, y_{2}$ and $\phi$  are
unchanged in the fragmentation process. The multidimensional distribution for both $c$ quarks 
(or both $\bar c$ antiquarks) is convoluted with fragmentation functions
simultaneously for each of the two quarks (or each of the two
antiquarks). As a result of the hadronization one obtains corresponding two-meson 
multidimensional distribution. In the last step experimental kinematical cuts on the distributions can be
imposed. Then the resulting distributions can be compared with
experimental ones. For numerical calculations here we apply often used in the case of heavy quarks, the Peterson fragmentation function \cite{Peterson}. We have shown in Ref.~\cite{MS2013-charmed-meson} that this scheme works very well in the case of inclusive $D^{0}$ meson spectra as well as for $D^{0}\bar{D^{0}}$ kinematical correlations.

%-----------------------
\section{Results}
%-----------------------

\label{sec:results}

%-------------------------------
\subsection{Parton level}
%-------------------------------

\begin{figure}[!h]
\begin{minipage}{0.47\textwidth}
 \centerline{\includegraphics[width=1.0\textwidth]{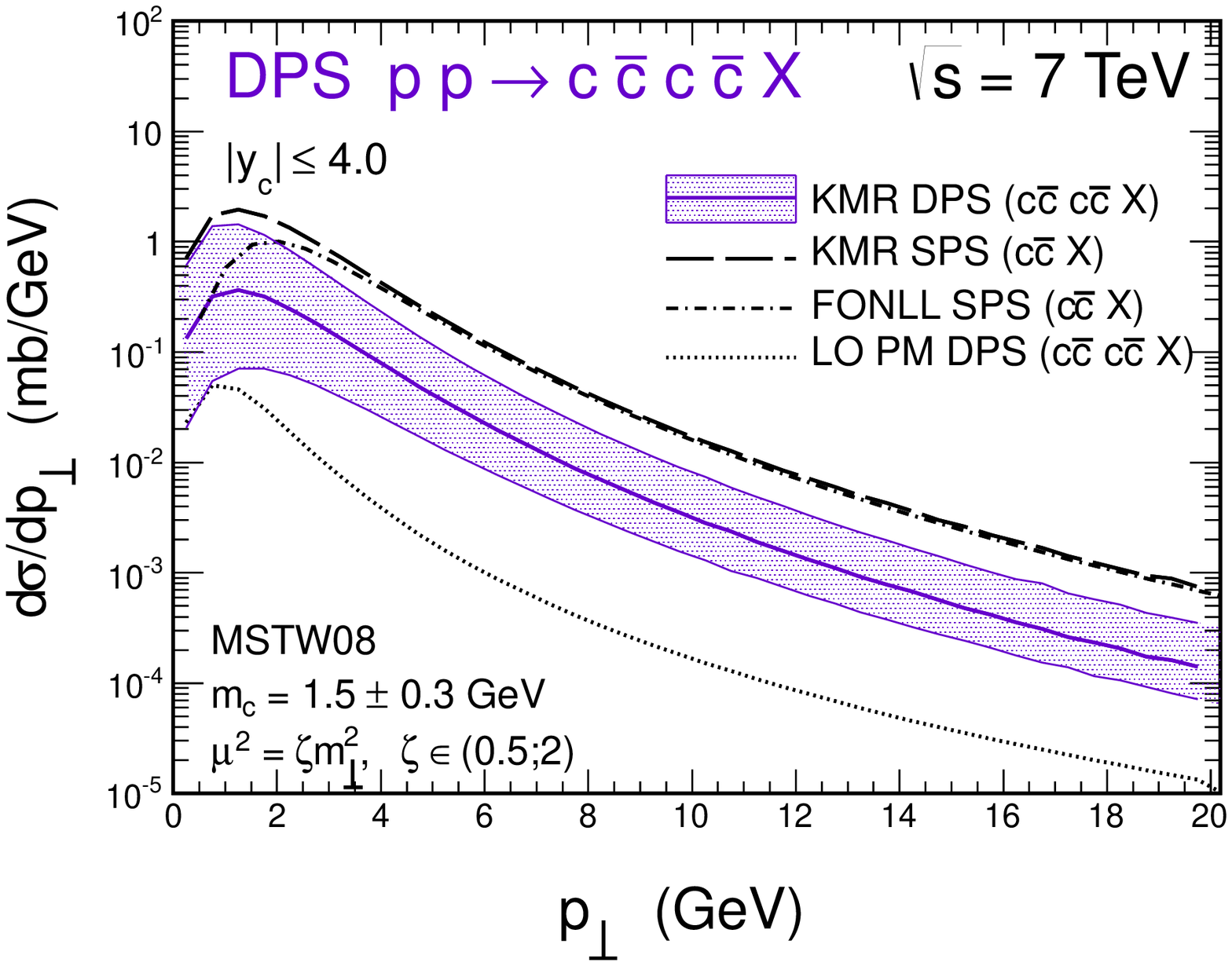}}
\end{minipage}
\hspace{0.5cm}
\begin{minipage}{0.47\textwidth}
 \centerline{\includegraphics[width=1.0\textwidth]{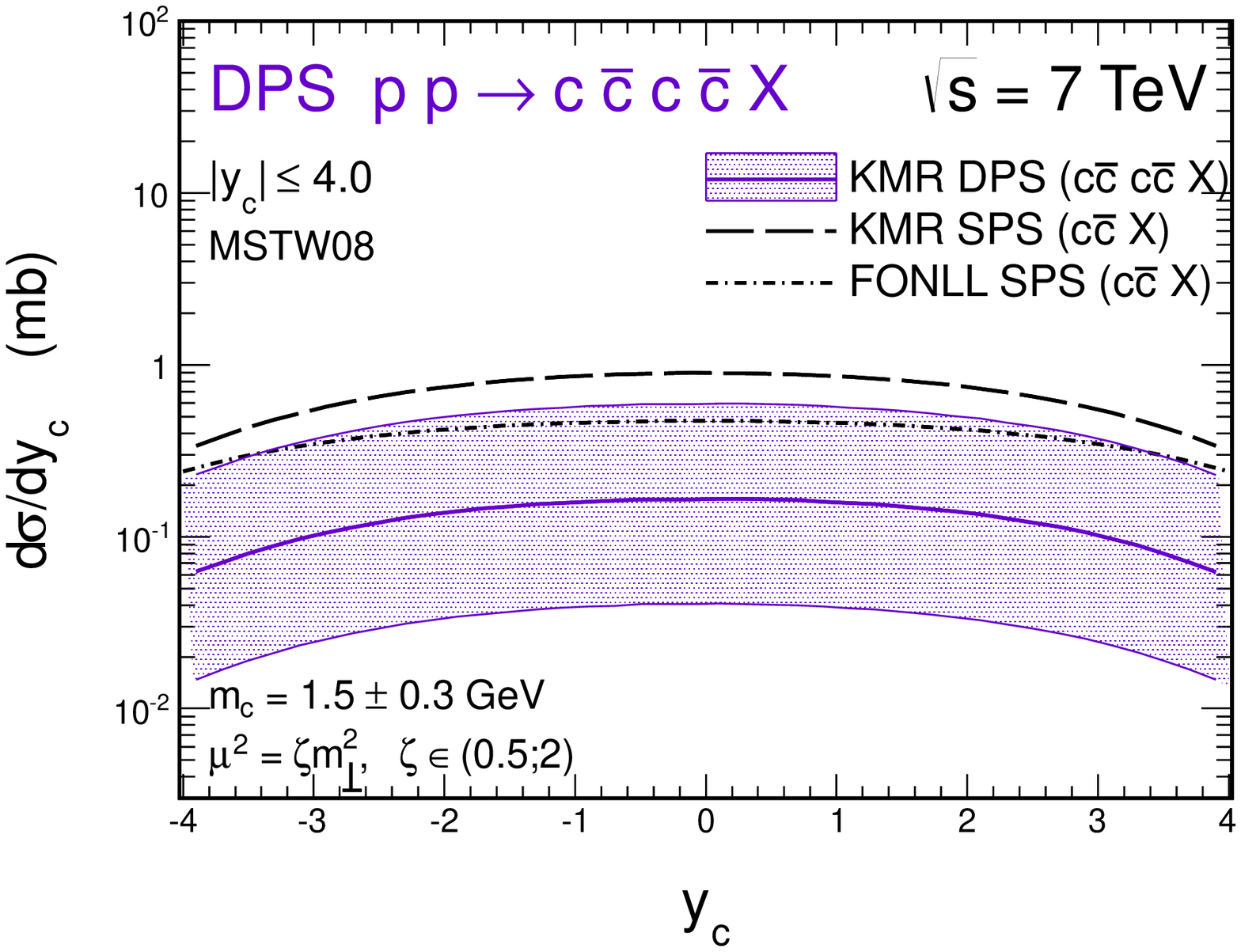}}
\end{minipage}
   \caption{
\small Transverse momentum (left) and rapidity (right)  of charm 
quarks from SPS $c \bar c$ (long-dashed line) and DPS $c \bar c c \bar c$ (solid line with shaded band) production. In this
calculation the KMR UGDF was used and the factorization scale and quark mass for the 
DPS contribution were varied as explained in the figure. For comparison 
LO collinear DPS distribution (dotted line) and FONLL SPS $c \bar c$ result
(dash-dotted line) are shown.
}
 \label{fig:DPS-KMR-uncertainties}
\end{figure}

We start from inclusive distributions of charm quarks (or antiquarks).
As discussed in Ref.~\cite{MS2013-charmed-meson} the
standard single-parton scattering contribution to $pp \to c \bar c X$ seems insufficient to 
describe inclusive spectra of charmed mesons as measured by 
the ATLAS, ALICE and LHCb collaborations \cite{ATLASincD,ALICEincD,LHCbincD}. 
The $c \bar c c \bar c$ production also contributes to 
the inclusive charm production. In Fig.~\ref{fig:DPS-KMR-uncertainties} 
we show such a contribution to transverse momentum distribution (left panel)
and rapidity distribution (right panel) together with theoretical uncertainty band.
In this calculation the Kimber-Martin-Ryskin (KMR) UGDF \cite{KMR} was used with the MSTW08 \cite{MSTW08} collinear gluon PDF. The solid line
corresponds to central value of our predictions.
The uncertainties are obtained by changing charm quark mass $m_c = 1.5\pm 0.3$ GeV which in general is not well known and by varying
renormalization and factorization scales $\mu^2 = \mu^2_{R} = \mu^2_{F} = \zeta m_{t}^2$, where $\zeta \in (0.5;2)$.
The shaded bands represent these both sources of uncertainties summed in quadrature.
As a reference point we plot contribution from standard single-scattering $c \bar c$ production, obtained
in the $k_t$-factorization approach (long-dashed line) as well as calculated with the help of FONLL code \cite{FONLL} (dash-dotted line). 
As can be seen both of these models are consistent and give very similar numerical results. It suggests that in the case of charm quark production
the $k_t$-factorization approach with the KMR UGDFs very well reproduces NLO corrections. These aspects of $c\bar{c}$ production
were discussed in more detail in Ref.~\cite{MS2013-charmed-meson}.
 
Since the DPS uncertainty band is very broad it becomes clear that 
this contribution is quite sizeable and must be included in the total
balance of charm quark (atniquark) production. For comparison we show
also DPS result obtained previously in Ref.~\cite{LMS2012} in the LO collinear approach. 
It is much smaller than the $k_t$-factorization result, especially at
larger transverse momenta. 

%-----------------------------------------------------------------------------
%------------------------------------------------------------------------------

\begin{figure}[!h]
\begin{minipage}{0.47\textwidth}
 \centerline{\includegraphics[width=1.0\textwidth]{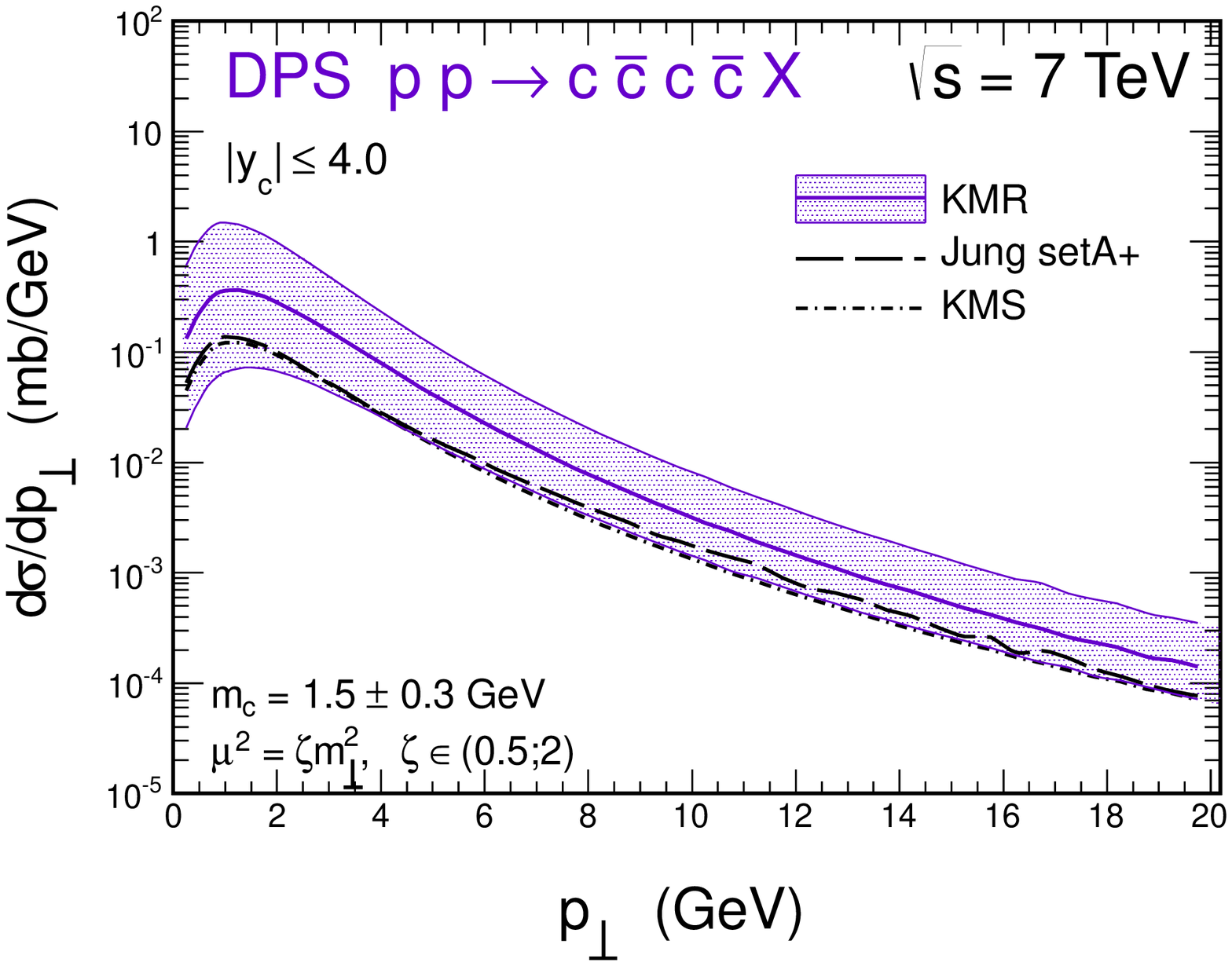}}
\end{minipage}
\hspace{0.5cm}
\begin{minipage}{0.47\textwidth}
 \centerline{\includegraphics[width=1.0\textwidth]{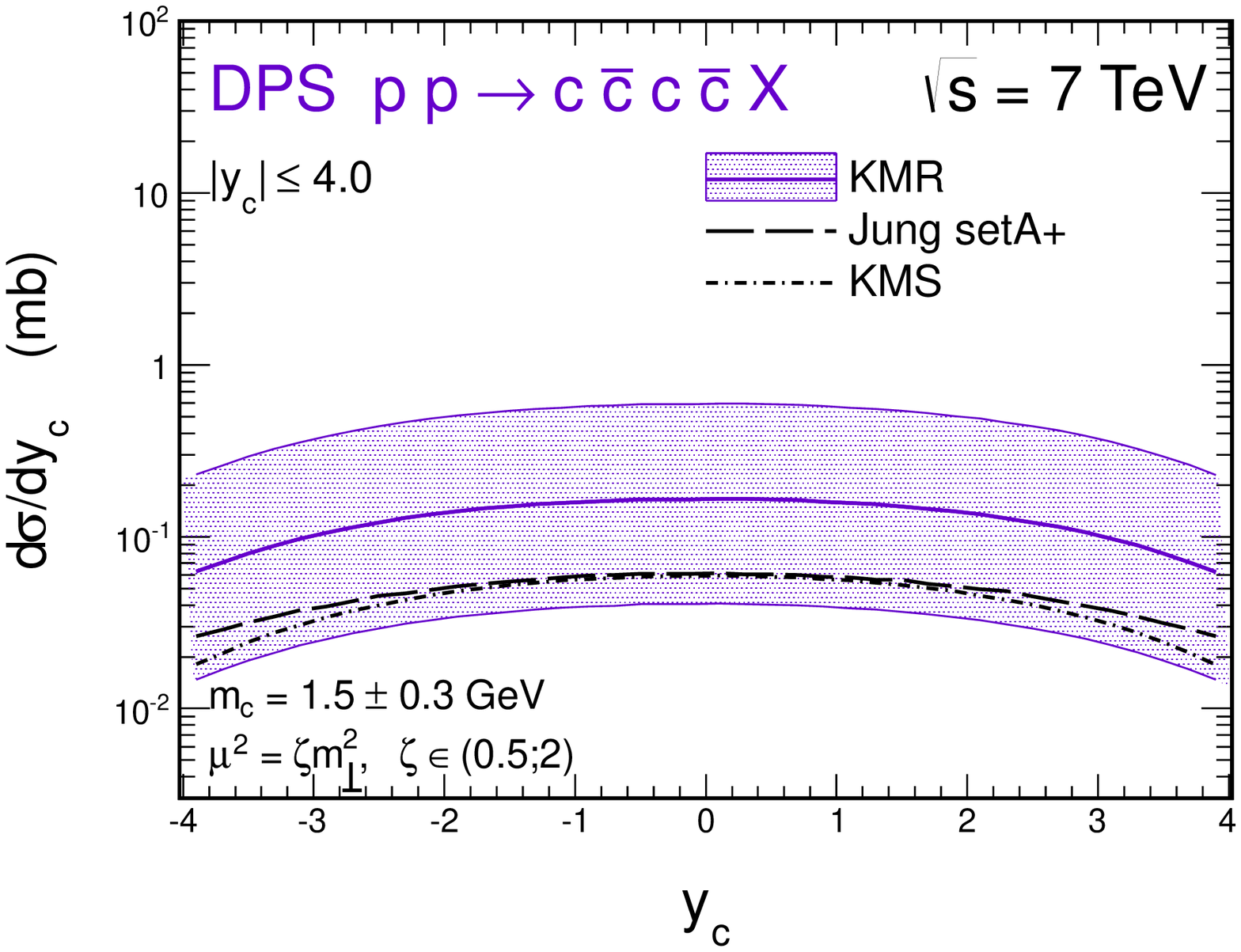}}
\end{minipage}
   \caption{
\small 
Transverse momentum (left) and rapidity (right) distributions of charm 
quarks produced in DPS $c \bar c c \bar c$ production for different
unintegrated gluon distributions.
}
 \label{fig:DPS-UGDFs}
\end{figure}

\begin{figure}[!h]
\begin{minipage}{0.47\textwidth}
 \centerline{\includegraphics[width=1.0\textwidth]{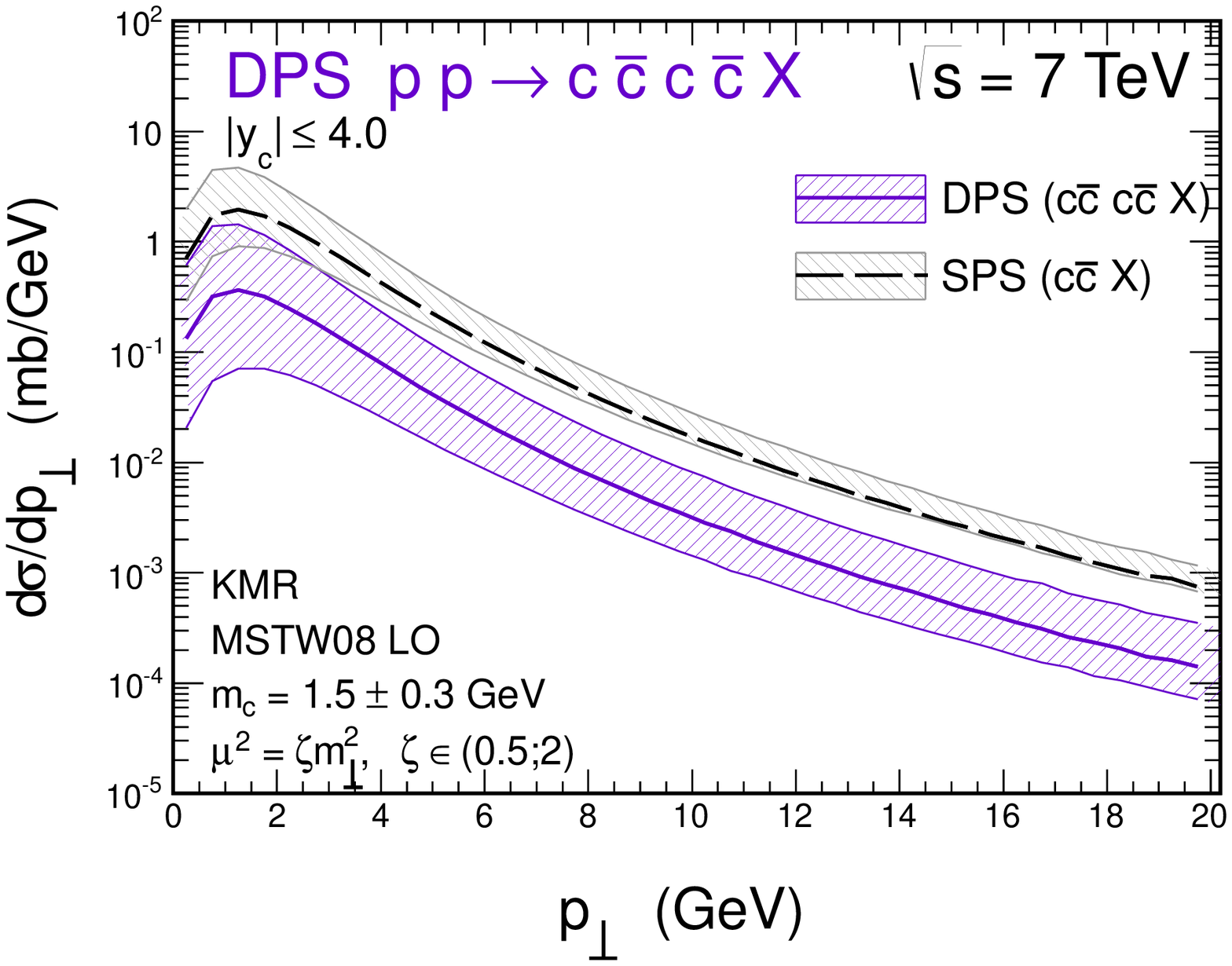}}
\end{minipage}
\hspace{0.5cm}
\begin{minipage}{0.47\textwidth}
 \centerline{\includegraphics[width=1.0\textwidth]{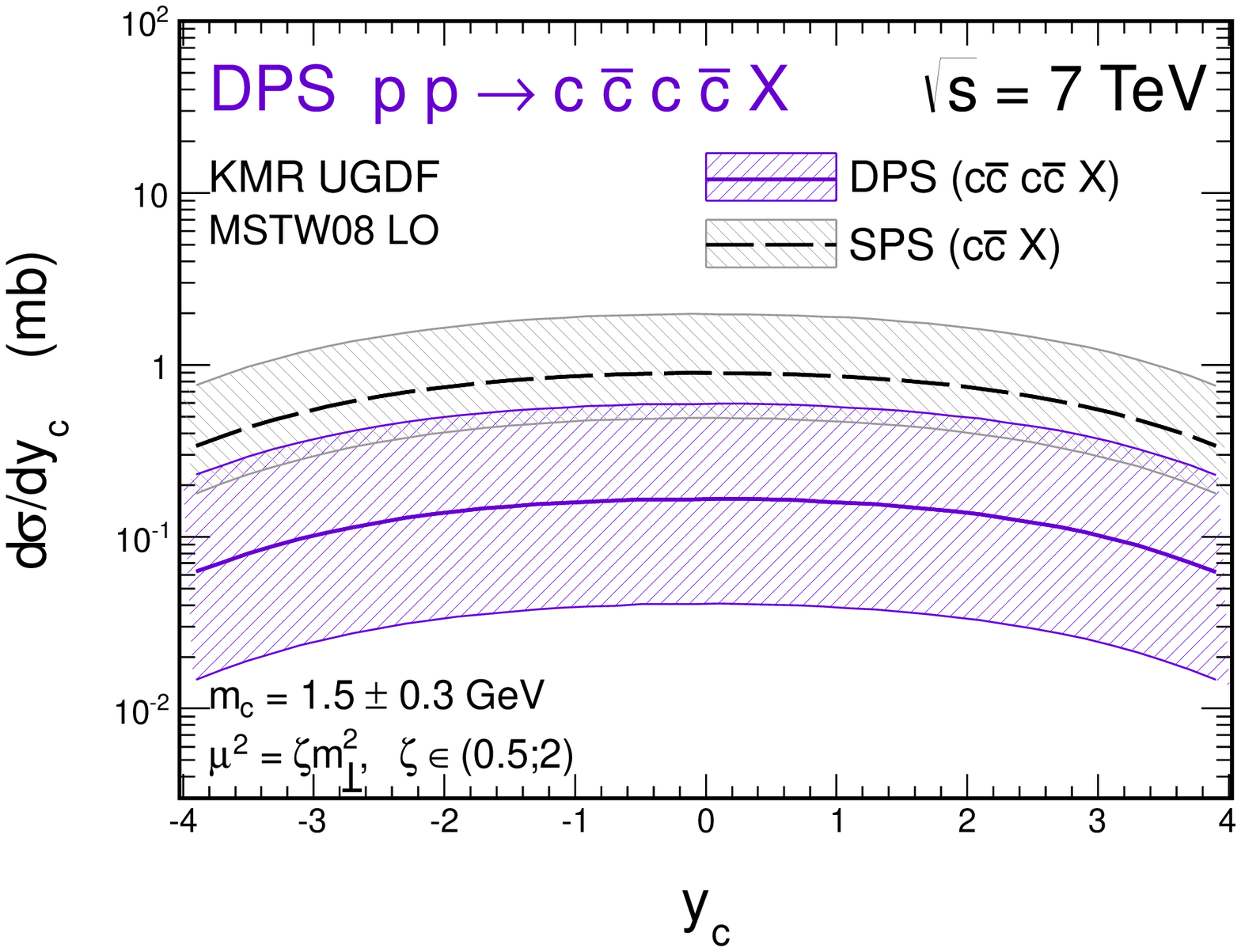}}
\end{minipage}
   \caption{
\small Comparison of the SPS $c \bar c$ and DPS $c \bar c c \bar c$
contributions to the inclusive charm quark (antiquark) production together with
theoretical uncertainties due to the choice of scales and those related with quark mass (summed in quadrature).
}
 \label{fig:DPS-3}
\end{figure}

In Fig.~\ref{fig:DPS-UGDFs} we compare DPS results for transverse momentum (left panel)
and rapidity (right panel) distributions obtained with different UGDFs from the
literature \cite{KMR,Jung,KMS}. The KMR UGDF gives the largest
cross section. Numerical results of DPS are more sensitive to the choice of UGDFs than in the case of SPS $c\bar c$ production, which can be understood by
different power of UGDFs in the cross section formula (fourth in DPS $c \bar c c \bar c$ versus second in SPS $c \bar c$). We use here also the KMS \cite{KMS} and Jung set$A+$ \cite{Jung} parametrizations. In turn, in Fig.~\ref{fig:DPS-3} we confront theoretical uncertainties of
SPS single pair ($c \bar c$) and DPS two-pair ($c \bar c c \bar c$)
production. Again uncertainty of the two pair production is much larger than
that for single pair production.

%-----------------------------------------------------------------------------
%------------------------------------------------------------------------------

%-----------------------------------------------------------------------------
%------------------------------------------------------------------------------

\begin{figure}[!h]
\begin{minipage}{0.47\textwidth}
 \centerline{\includegraphics[width=1.0\textwidth]{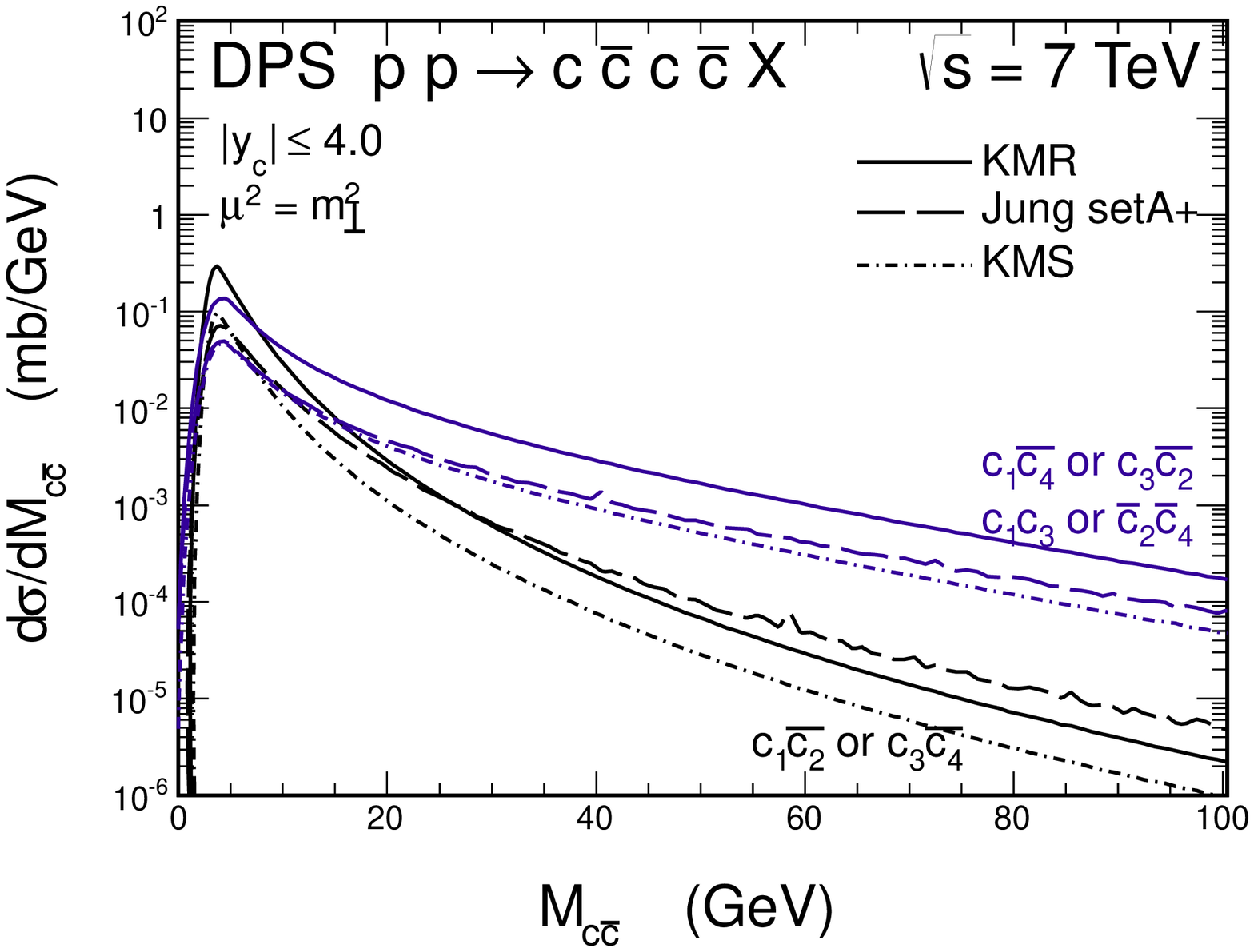}}
\end{minipage}
\hspace{0.5cm}
\begin{minipage}{0.47\textwidth}
 \centerline{\includegraphics[width=1.0\textwidth]{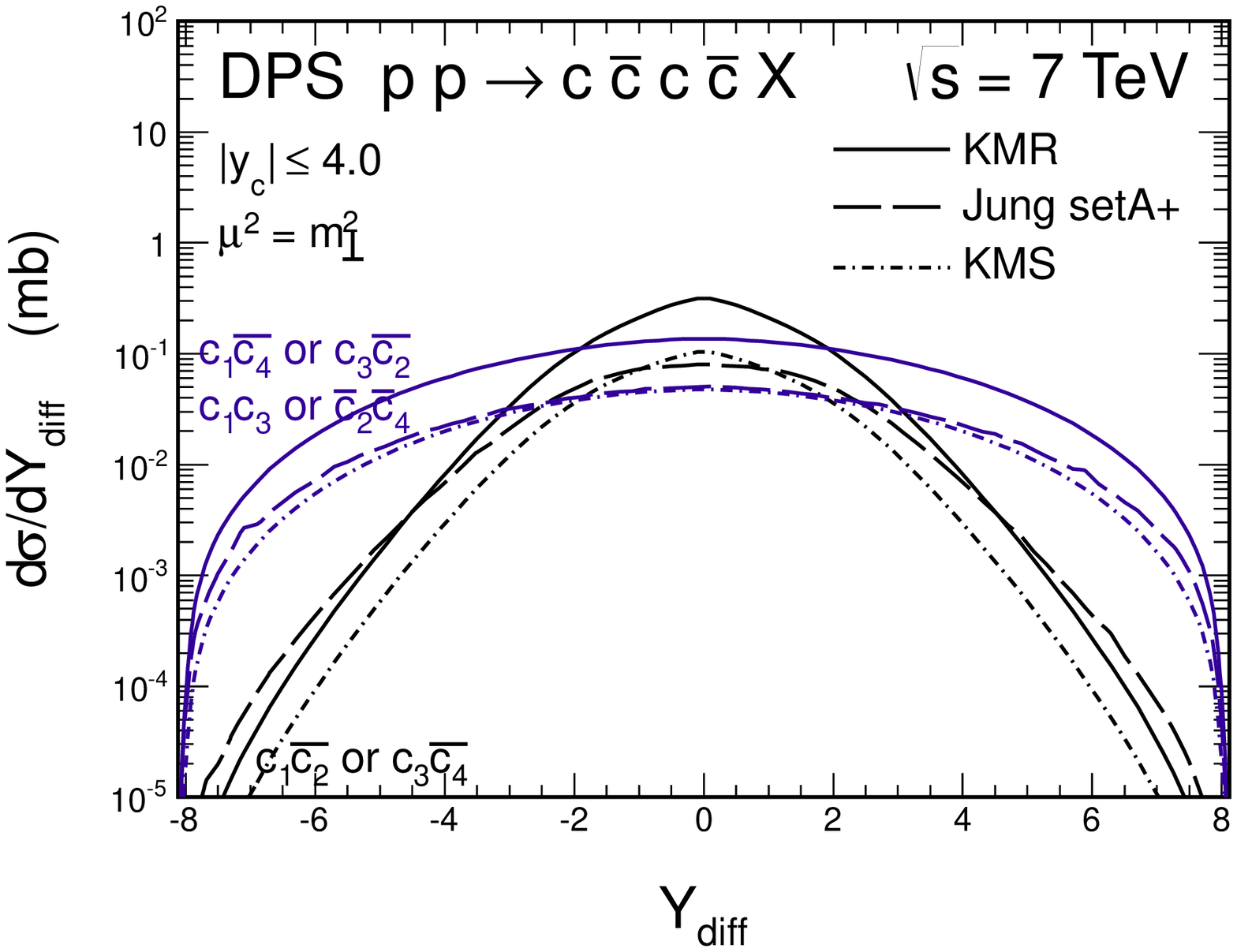}}
\end{minipage}
   \caption{
\small Distribution in the invariant mass of quark/atiquark $M_{c\bar c}$ (left) and distribution in the rapidity distance between quarks/antiquarks $Y_{diff}$ (right) from the
same ($c_1\bar c_2$ or $c_3\bar c_4$) and from different scatterings ($c_1\bar c_4$ or $c_3\bar c_2$ or $c_1 c_3$ or $\bar c_2\bar c_4$), calculated with different UGDFs.
}
\label{fig:correlations_kt_factorization_1}
\end{figure}

In Ref.~\cite{LMS2012} we have proposed several correlation distributions
to be studied in order to identify the DPS effects.
Here we present the same distributions as in Ref.~\cite{LMS2012} but within the
$k_t$-factorization approach. In
Fig.~\ref{fig:correlations_kt_factorization_1} we show distributions
in invariant mass $M_{c\bar c}$ (left panel) and rapidity difference of quarks/antiquarks $Y_{diff} = y_c - y_{\bar c}$ (right panel) from the
same scattering ($c_1\bar c_2$ or $c_3\bar c_4$) and from different scatterings ($c_1\bar c_4$ or $c_3\bar c_2$ or $c_1 c_3$ or $\bar c_2\bar c_4$) for various UGDFs specified
in the figure. The shapes of distributions in the figure are almost 
identical as their counterparts obtained in LO collinear
approach in Ref.~\cite{LMS2012}. 

%-----------------------------------------------------------------------------
%------------------------------------------------------------------------------

\begin{figure}[!h]
\begin{minipage}{0.47\textwidth}
 \centerline{\includegraphics[width=1.0\textwidth]{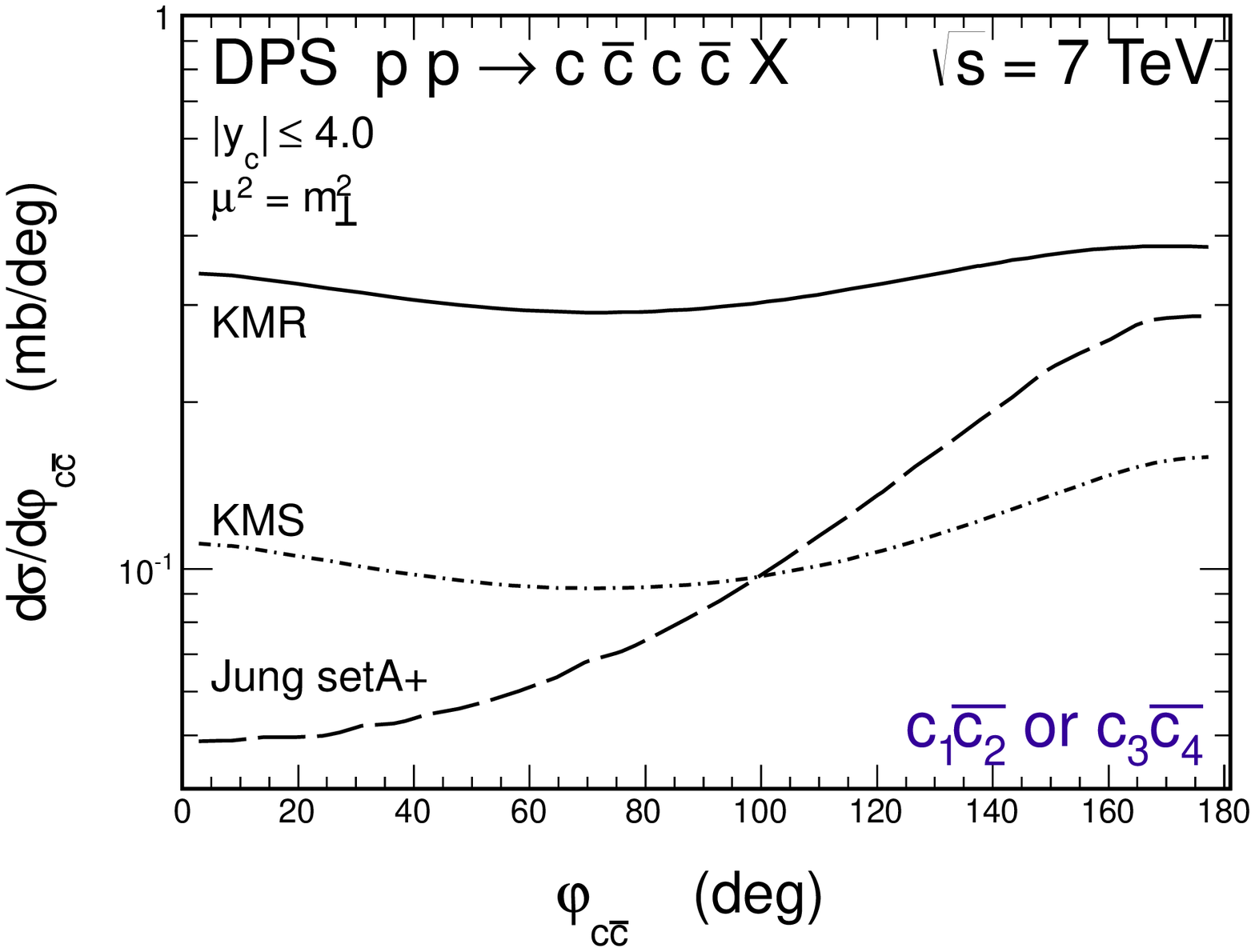}}
\end{minipage}
\hspace{0.5cm}
\begin{minipage}{0.47\textwidth}
 \centerline{\includegraphics[width=1.0\textwidth]{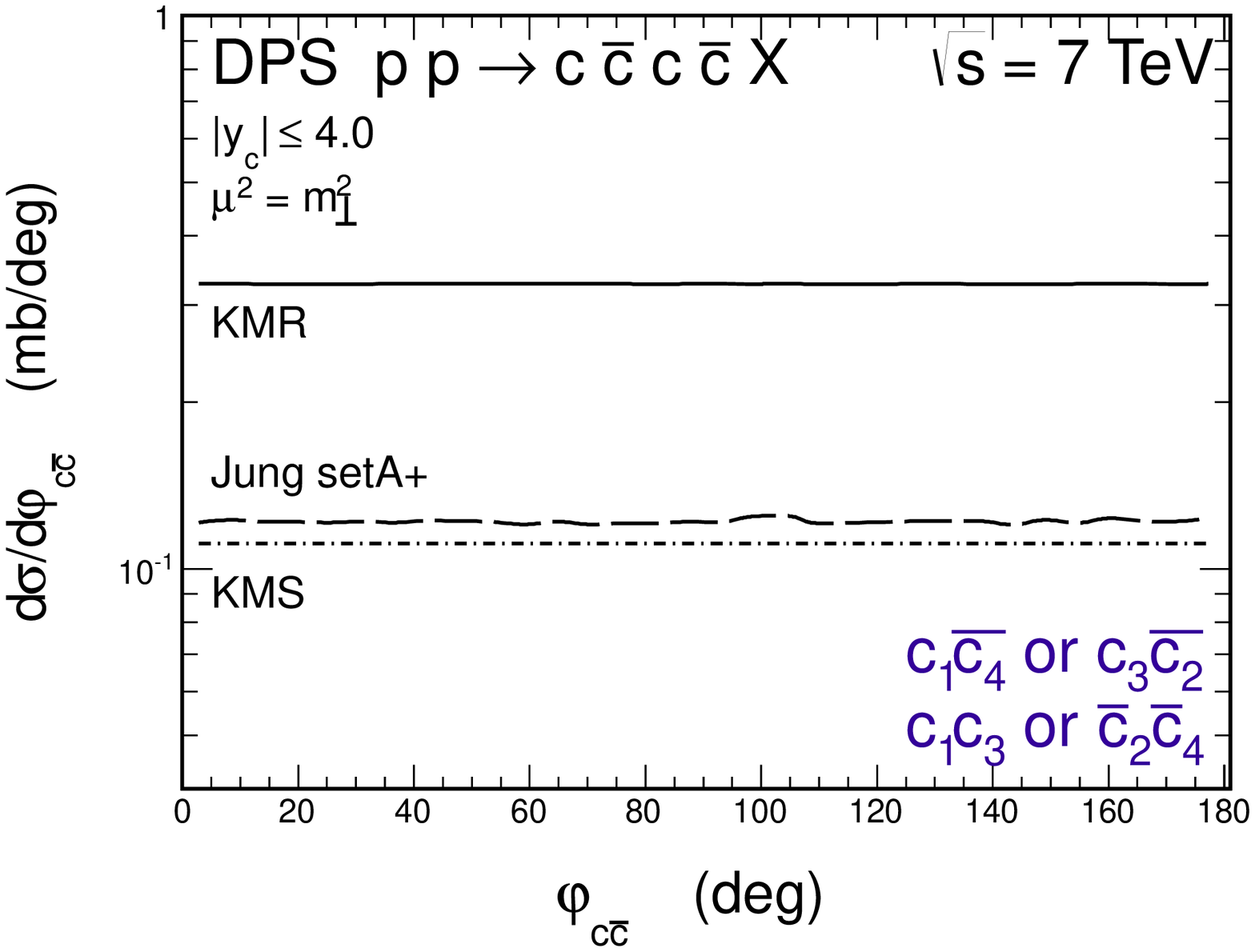}}
\end{minipage}
   \caption{
\small Distribution in azimuthal angle $\varphi_{c\bar c}$ between quarks/antiquarks from
the same scattering (left) and from different scatterings (right), calculated with different UGDFs. 
}
\label{fig:correlations_kt_factorization_2}
\end{figure}

In Fig.~\ref{fig:correlations_kt_factorization_2} we show distributions
in azimuthal angle difference between quarks/antiquarks $\varphi_{c\bar c}$ from the same
and from different scatterings. While in the case of the same
scattering distribution strongly depends on the choice of UGDF the
quarks/antiquarks from different scattering are not correlated which
is inherent property of the simple factorized model. Our distinguishing of scatterings can
be done only in the model calculation. Experimentally one observes both
types together after hadronization which naturally may bring additional
decorrelation.

%-----------------------------------------------------------------------------
%------------------------------------------------------------------------------

Finally, we present distribution in transverse momentum of the pair
of quarks $p_{\perp}^{c\bar c}$. In LO collinear approach the distribution for
emission in the same scattering is very different from the case
of emissions from different scatterings \cite{LMS2012}. The picture in 
the $k_t$-factorization approach is, however, very different. The respective 
distributions for the same and different scatterings are rather similar. 
This means that transverse momentum of the pair may not be the best
quantity to be used in order to identify the DPS effects.

%-----------------------------------------------------------------------------
\begin{figure}[!h]
\begin{minipage}{0.47\textwidth}
 \centerline{\includegraphics[width=1.0\textwidth]{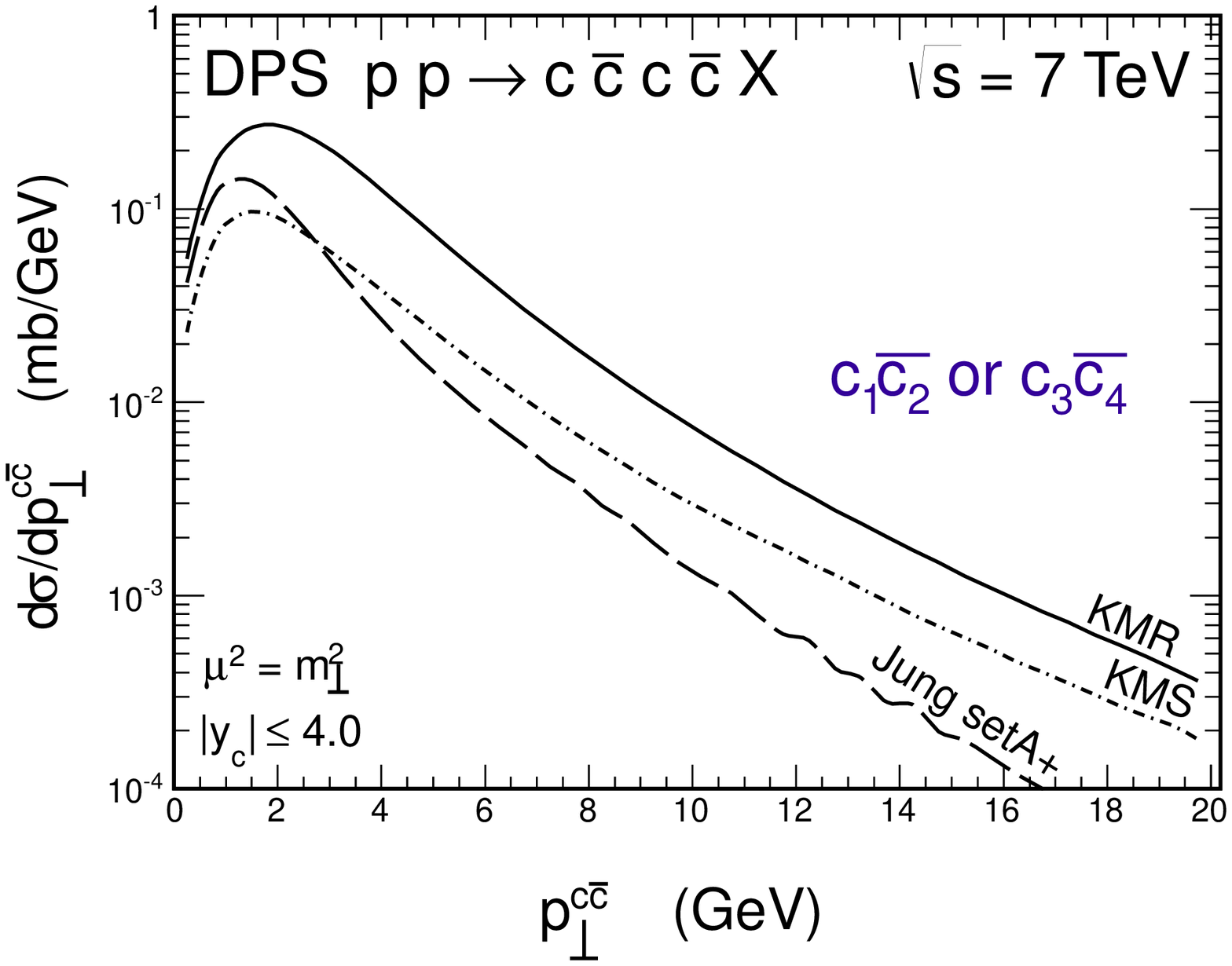}}
\end{minipage}
\hspace{0.5cm}
\begin{minipage}{0.47\textwidth}
 \centerline{\includegraphics[width=1.0\textwidth]{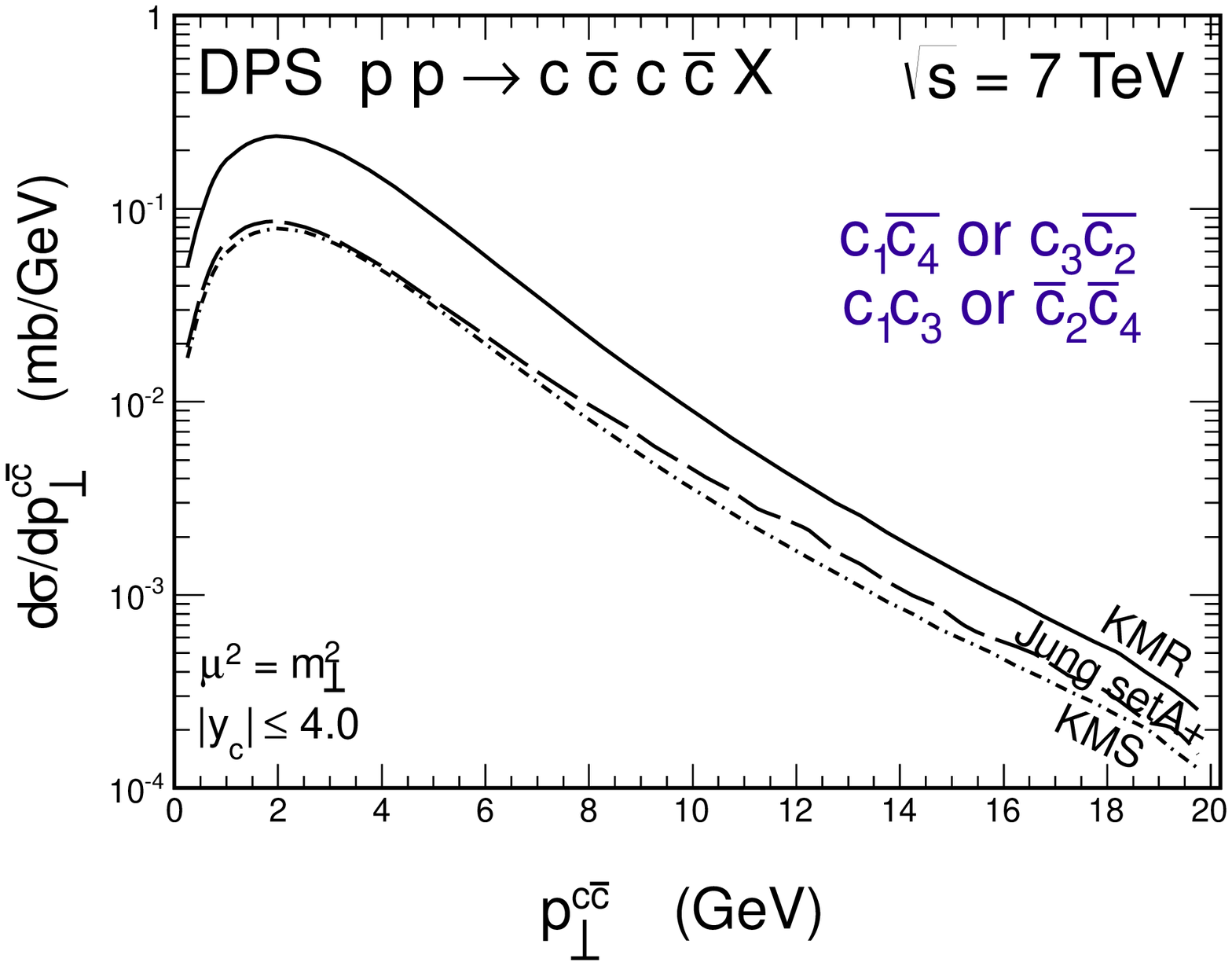}}
\end{minipage}
   \caption{
\small Distribution in transverse momentum of the quark/antiquark
pair $p_{\perp}^{c\bar c}$ from the same scattering (left) and from different scatterings
(right), calculated with different UGDFs. 
}
\label{fig:correlations_kt_factorization_3}
\end{figure}
%------------------------------------------------------------------------------

%-----------------------------
\subsection{Meson level}
%-----------------------------

Production of two pairs of $c \bar c$ on the partonic level leads to 
the situations that very often two mesons, both containing $c$ quarks or/and
both containing $\bar c$ antiquarks, are produced on the hadronic level
in one event. Therefore the presence of two such mesons may be considered
as a signal of production of $c \bar c c \bar c$ on the partonic level.
Recently, the LHCb collaboration performed a first measurement of $D_i D_j + \bar{D_i}\bar{D_j}$ production
in the fiducial range of the detector acceptance $2 < y_D < 4$ and $3 < p^{D}_{\perp} < 12$ GeV \cite{LHCb-DPS-2012}.
As described in Section \ref{sec:theory} we have prepared a code
which keeps track of full kinematical information about each of both quarks
or both antiquarks and similar information about both mesons.
Such a multidimensional map is used then to impose adequate experimental cuts.

%------------------------------------------------------------------------------
\begin{table}[tb]%
\caption{Total cross sections for meson-meson pair production for three different UGDFs.}
\newcolumntype{Z}{>{\centering\arraybackslash}X}
\label{table}
\centering %
\begin{tabularx}{16.5cm}{ZZZZZZZZZZ}
\toprule[0.1em] %
\\[-2.4ex] 
%\\[1.0ex]

 \multirow{3}*{Mode} & &  & \multicolumn{7}{c}{$\sigma_{tot}^{THEORY} \;\;$[nb]} \\ [+0.4ex]
                     & \multicolumn{2}{c}{$\sigma_{tot}^{EXP} \;\;$[nb]}  & \multicolumn{3}{c}{KMR $^{+}_{-}(\mu)$ $^{+}_{-}(m_{c})$} & \multicolumn{2}{c}{Jung setA$+$} & \multicolumn{2}{c}{KMS} \\[+0.4ex] 
                     &                                             &  & \multicolumn{2}{c}{$\varepsilon_{c} = 0.05$}  &  $\varepsilon_{c} = 0.02$  & $\varepsilon_{c} = 0.05$ & $\varepsilon_{c} = 0.02$ & $\varepsilon_{c} = 0.05$ & $\varepsilon_{c} = 0.02$ \\[+0.4ex]

\toprule[0.1em]\\[-2.4ex] 

 $D^{0}D^{0}$          & \multicolumn{2}{c}{$690\pm40\pm70$} & \multicolumn{2}{c}{$265$ $^{+140}_{-77}$ $^{+157}_{-94}$} & 400  & 120 & 175  & 84 & 126  \\ [+0.4ex]
 $D^{0}D^{+}$          & \multicolumn{2}{c}{$520\pm80\pm70$} & \multicolumn{2}{c}{$212$ $^{+112}_{-62}$ $^{+126}_{-75}$} & 319  & 96  & 140  & 67 & 100 \\ [+0.4ex]
 $D^{0}D^{+}_{S}$      & \multicolumn{2}{c}{$270\pm50\pm40$} & \multicolumn{2}{c}{$75$  $^{+40}_{-22}$ $^{+45}_{-27}$}  & 113  & 34  & 50   & 24  & 36 \\  [+0.4ex]
 $D^{+}D^{+}$          & \multicolumn{2}{c}{$80\pm10\pm10$}  & \multicolumn{2}{c}{$42$  $^{+23}_{-13}$ $^{+26}_{-15}$}   & 64   & 19  & 28   & 13  & 20 \\[+0.4ex] 
 $D^{+}D^{+}_{S}$      & \multicolumn{2}{c}{$70\pm15\pm10$}  & \multicolumn{2}{c}{$30$  $^{+16}_{-9}$ $^{+18}_{-11}$}     & 45   & 14  & 20   & 10  & 14 \\[+0.4ex]
 $D^{+}_{S}D^{+}_{S}$  & \multicolumn{2}{c}{$-$}             & \multicolumn{2}{c}{$11$  $^{+5}_{-3}$ $^{+6}_{-4}$}      & 16   & 5   & 7    & 3   & 5 \\[+0.4ex]
\hline
\bottomrule[0.1em]

\end{tabularx}

\end{table}
%-----------------------------------------------------------------------------

In Table 1 we have collected DPS cross sections for different pairs of mesons relevant for considered kinematics
obtained with different unintegrated gluon distributions.
As was shown in Ref.~\cite{MS2013-charmed-meson} theoretical predictions for production of charmed meson pairs in the LHCb kinematics are
very sensitive to the value of $\varepsilon_{c}$ parameter in the Peterson fragmentation function. There, rather 
harder functions (with smaller $\varepsilon_{c}$) are suggested for better description of experimental data,
which is also in agreement with observations made in the FONLL framework \cite{Cacciari,Cacciari-RHIC}.
Therefore we present results for two different values of the $\varepsilon_{c}$ parameter.
Here we have added together cross sections for 
charge conjugated channels: $\sigma_{D_i D_j} + \sigma_{\bar D_i \bar D_j}$.
The calculated cross sections are somewhat smaller than the
experimental ones. Only the upper limit of our predictions with the Kimber-Martin-Ryskin UGDF and with $\varepsilon_{c} = 0.02$ in the Peterson fragmentation function gives results which
are close to the experimental data, taken the uncertainties on
the choice of the factorization/renormalization scale and on the charm quark mass.

\begin{figure}[!h]
\begin{minipage}{0.47\textwidth}
 \centerline{\includegraphics[width=1.0\textwidth]{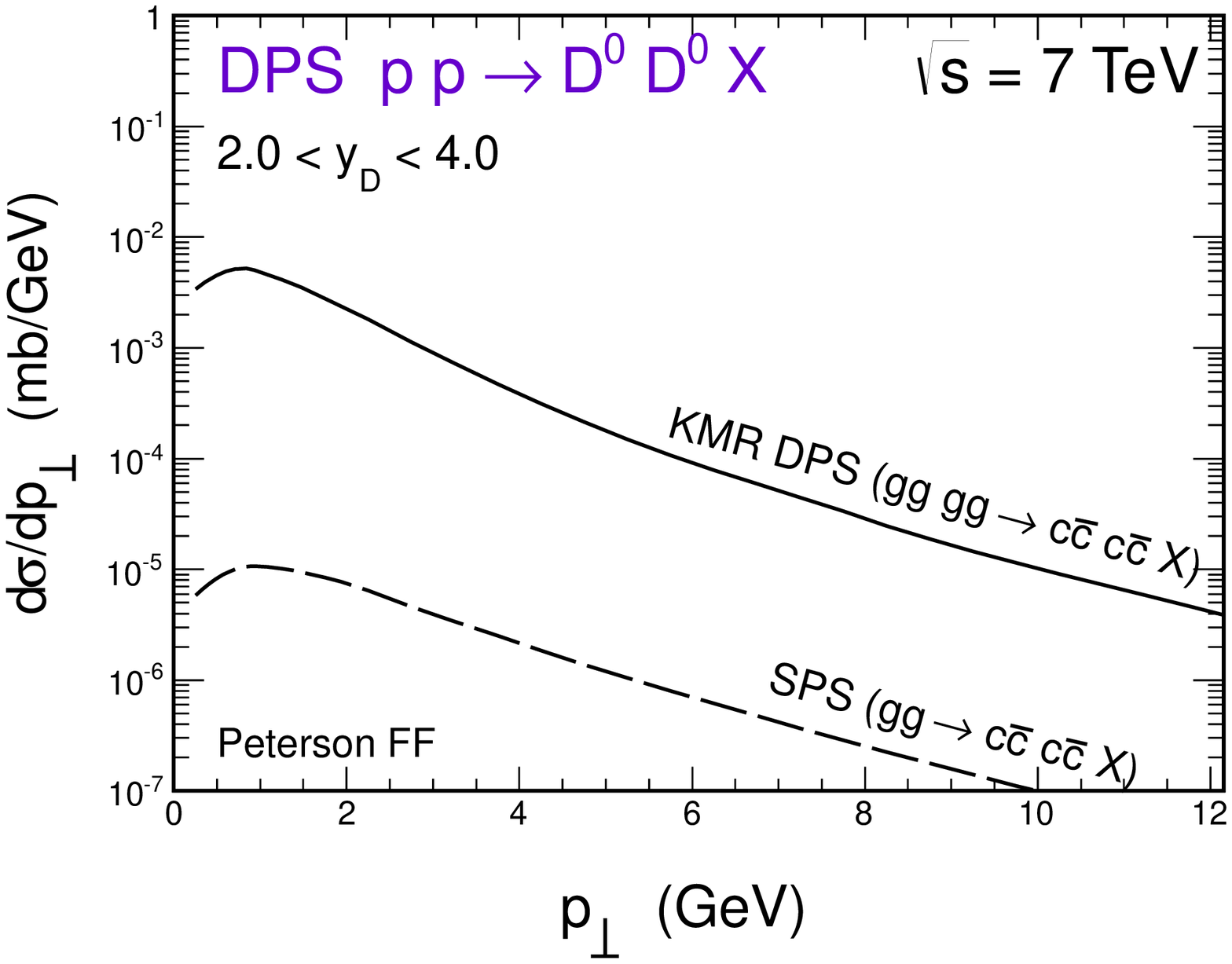}}
\end{minipage}
\hspace{0.5cm}
\begin{minipage}{0.47\textwidth}
 \centerline{\includegraphics[width=1.0\textwidth]{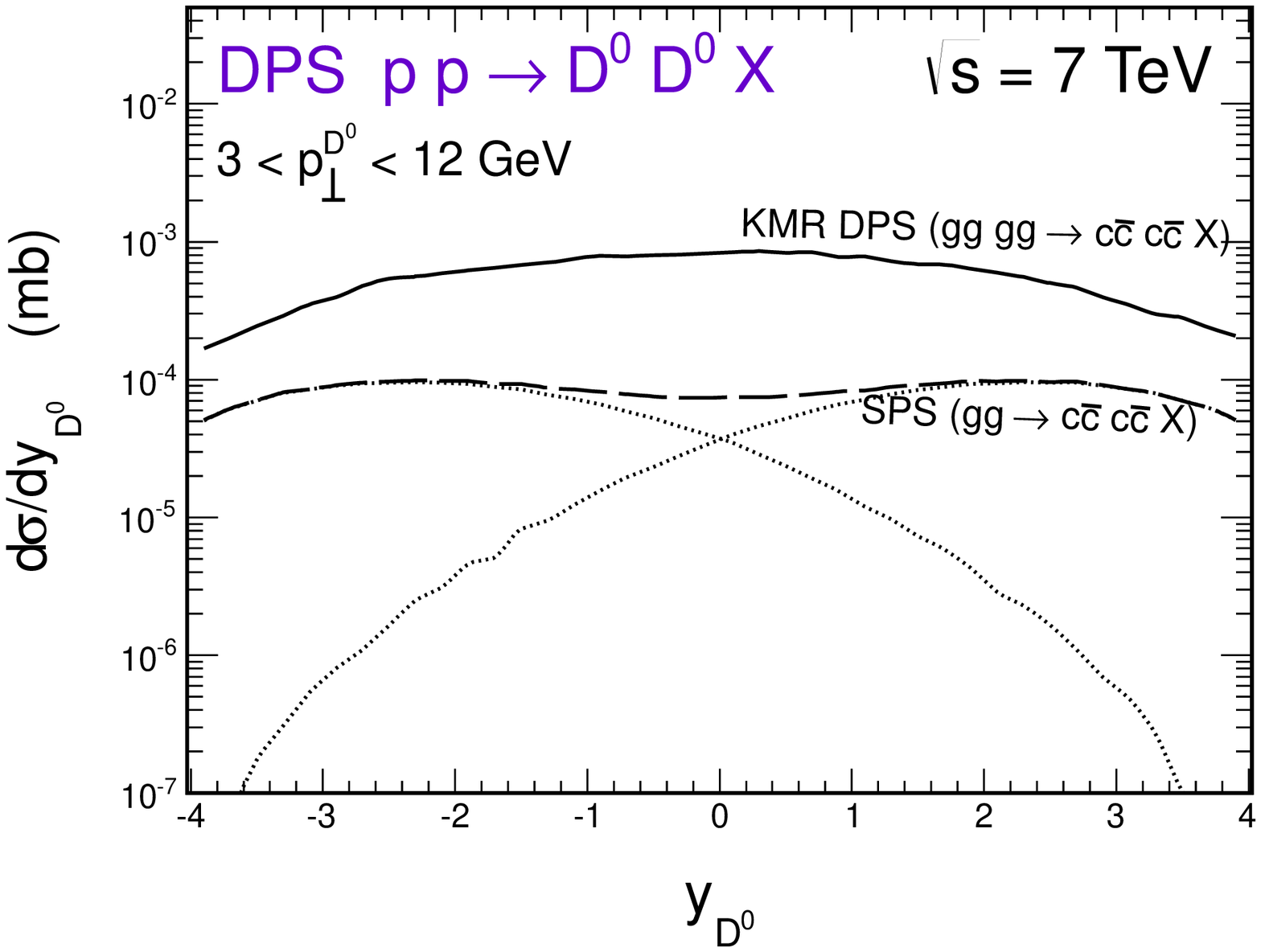}}
\end{minipage}
   \caption{
\small Distributions in  transverse momentum (left panel) and rapidity (right panel) of single $D^{0}$ meson from the $D^0 D^0$ pair events. The solid lines corresspond to DPS mechanism and the long-dashed lines represent contributions from SPS production of $D^{0}D^{0}$ pairs. Here we impose kinematical cuts adequate for the LHCb kinematics.}
\label{fig:SPSvsDPS-lhcb}
\end{figure}

\begin{figure}[!h]
\begin{minipage}{0.35\textwidth}
 \centerline{\includegraphics[width=1.0\textwidth]{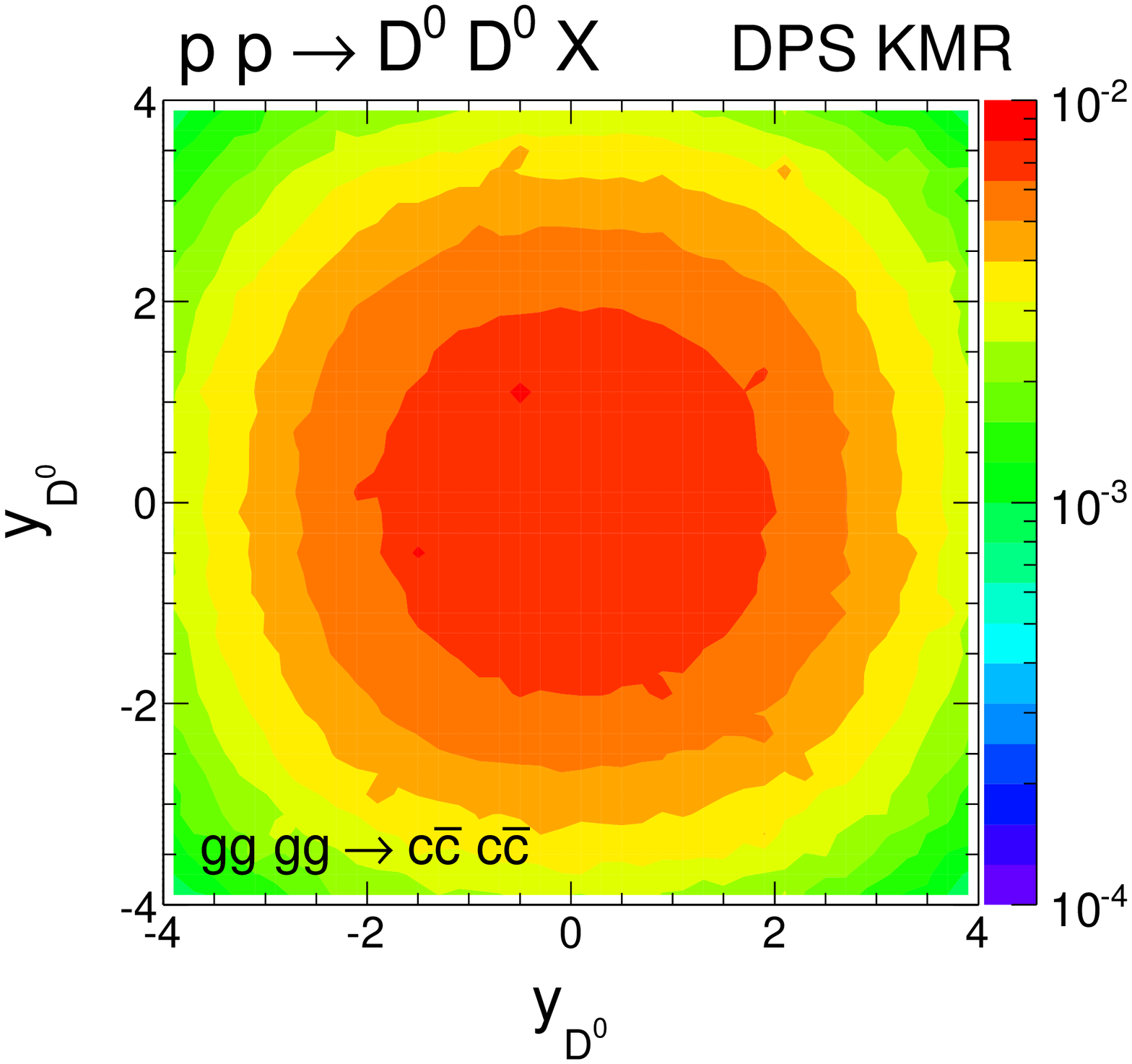}}
\end{minipage}
\hspace{0.5cm}
\begin{minipage}{0.35\textwidth}
 \centerline{\includegraphics[width=1.0\textwidth]{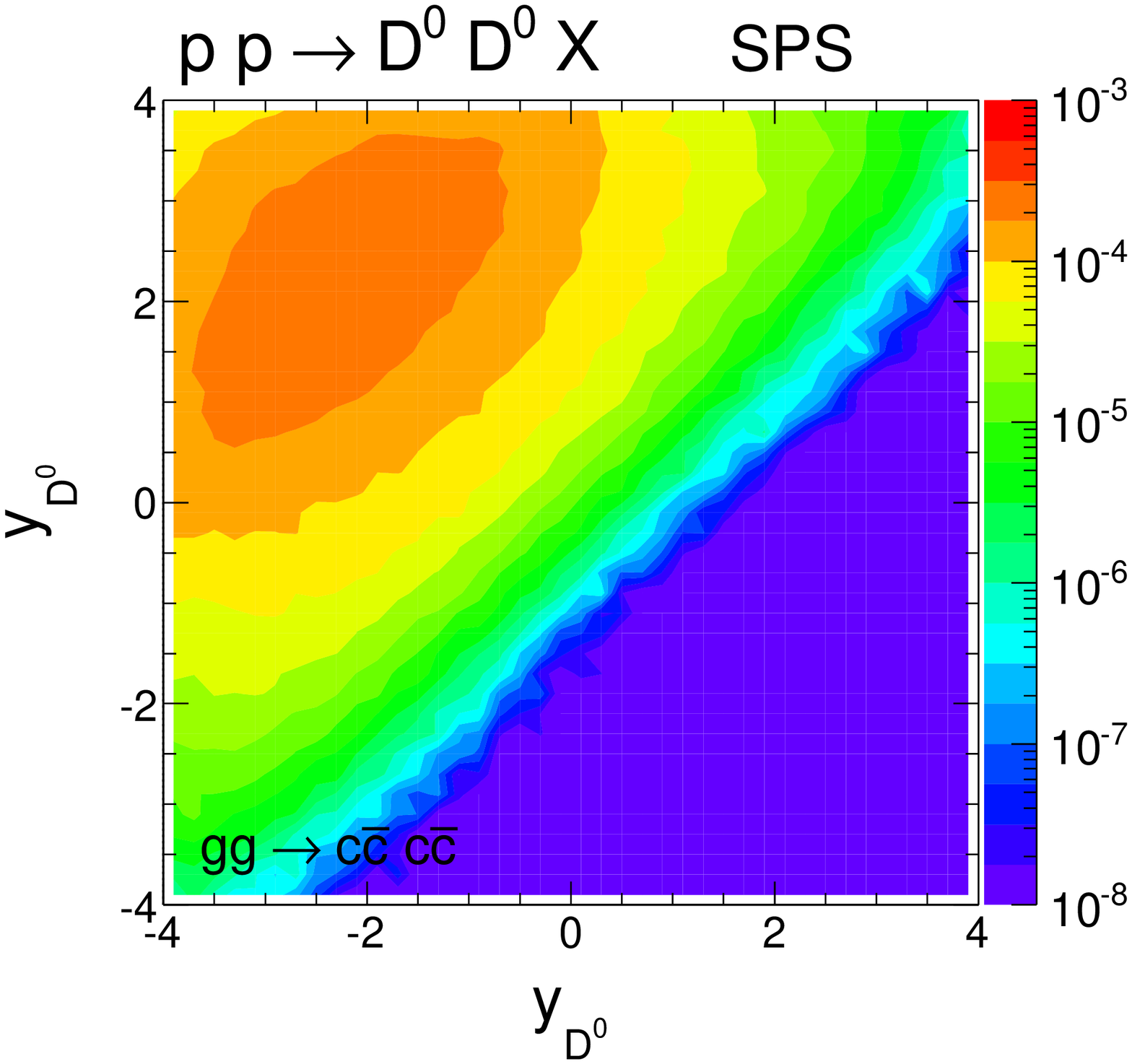}}
\end{minipage}
   \caption{
\small Two-dimensional distributions in rapidity of one $D^0$ meson and rapidity of the second $D^{0}$ meson for DPS (left) and SPS (right).
}
\label{fig:SPSvsDPS-lhcb-2dim}
\end{figure}

So far we have considered only DPS contribution to $D_iD_j$ (or $\bar{D_i}\bar{D_j}$) production. In Fig.~\ref{fig:SPSvsDPS-lhcb}
we show in addition corresponding SPS contribution. The SPS contribution to the transverse momentum distribution (left panel) is more than two
orders of magnitude smaller than the DPS one. For the rapidity distribution (right panel)
the difference is only one order of magnitude. This effect is slightly unintuitive.
However, it can be understood by a comparison of the two-dimensional distributions
in rapidity of one and the second $D$ meson for DPS and SPS production (see Fig.~\ref{fig:SPSvsDPS-lhcb-2dim}).
In the case of DPS the two mesons are not correlated (in this plane), in contrast to the SPS mechanism, where
they are strongly anticorrelated. When one meson is produced in forward rapidity
region the second is preferentially produced in backward rapidity region, or
vice versa. One can also conclude that in the case of $D_iD_j$ ($\bar D_{i}\bar D_{j}$) pair production,
the specific LHCb kinematical range leads to a dumping of the SPS cross section. 
The requirement that both $D$ mesons have to reach the detector makes
the SPS contribution almost negligible. Quite different conclusions can be drawn in the case of inclusive $D$ meson measurements.

In the present paper we have calculated SPS $c \bar c c \bar c$ contribution in the
high-energy approximation which may not be the best approximation
for the LHCb kinematics where the distance between both $c$ or both
$\bar c$ is rather small. Therefore, to drawn definite conclusions,
future studies of the $p p \to c \bar c c \bar c X$ process are needed and they must include a
complete set of diagrams for the SPS $c \bar c c \bar c$ mechanism.
Furthermore, if the improved calculations of SPS mechanism will not provide somewhat
better description of the total cross sections measured by LHCb, one has to look for
other mechanisms which can contribute and fill predicted missing strength.

%-----------------------------------------------------------------------------
%------------------------------------------------------------------------------

%-----------------------------------------------------------------------------
%------------------------------------------------------------------------------

The LHCb collaboration presented also several differential distributions for
the simultaneous production of two $DD$ and $\bar D \bar D$ mesons.
Here we consider only examples for $D^0 D^0$ (identical to 
$\bar D^0 \bar D^0$) channel.

\begin{figure}[!h]
\begin{minipage}{0.47\textwidth}
 \centerline{\includegraphics[width=1.0\textwidth]{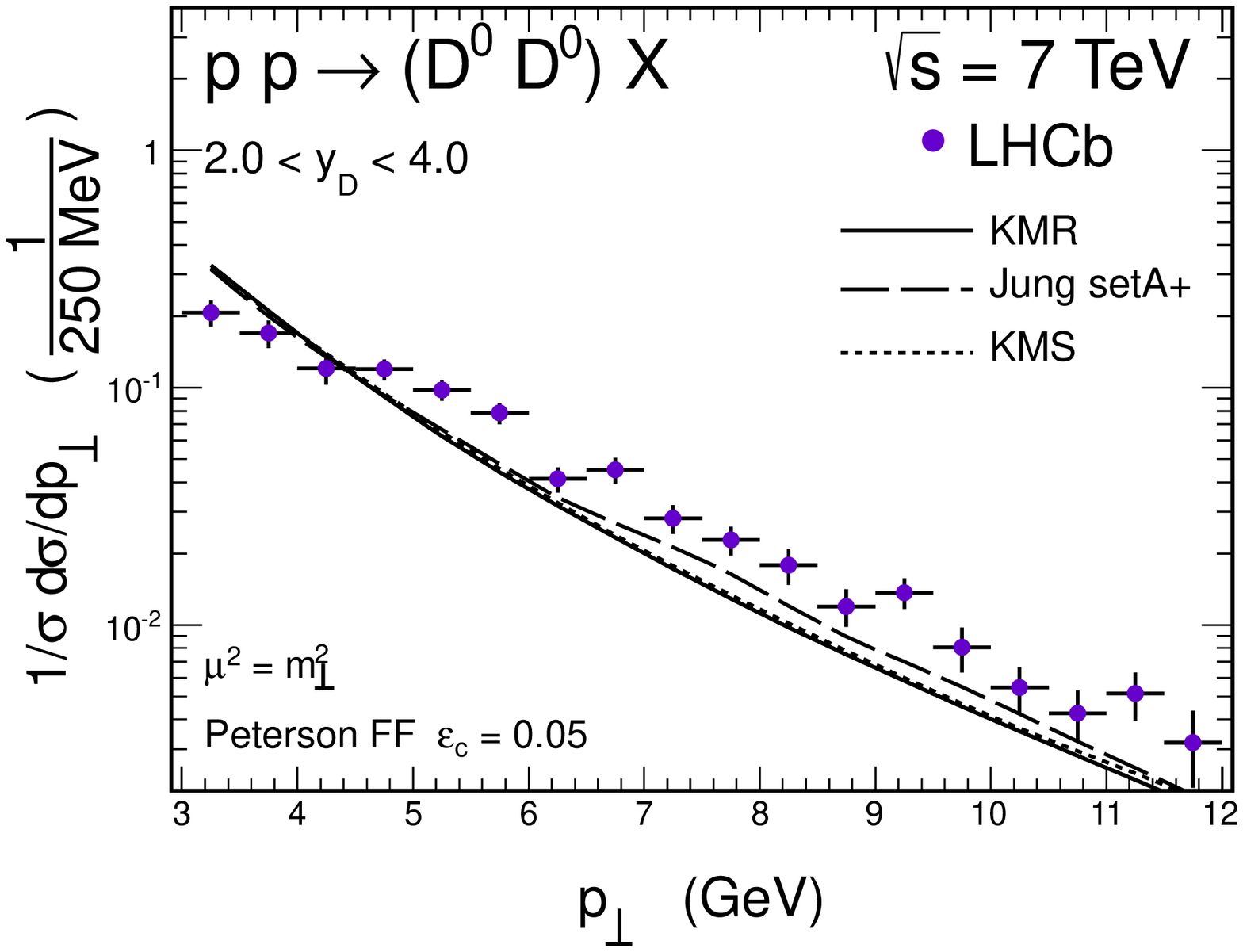}}
\end{minipage}
\hspace{0.5cm}
\begin{minipage}{0.47\textwidth}
 \centerline{\includegraphics[width=1.0\textwidth]{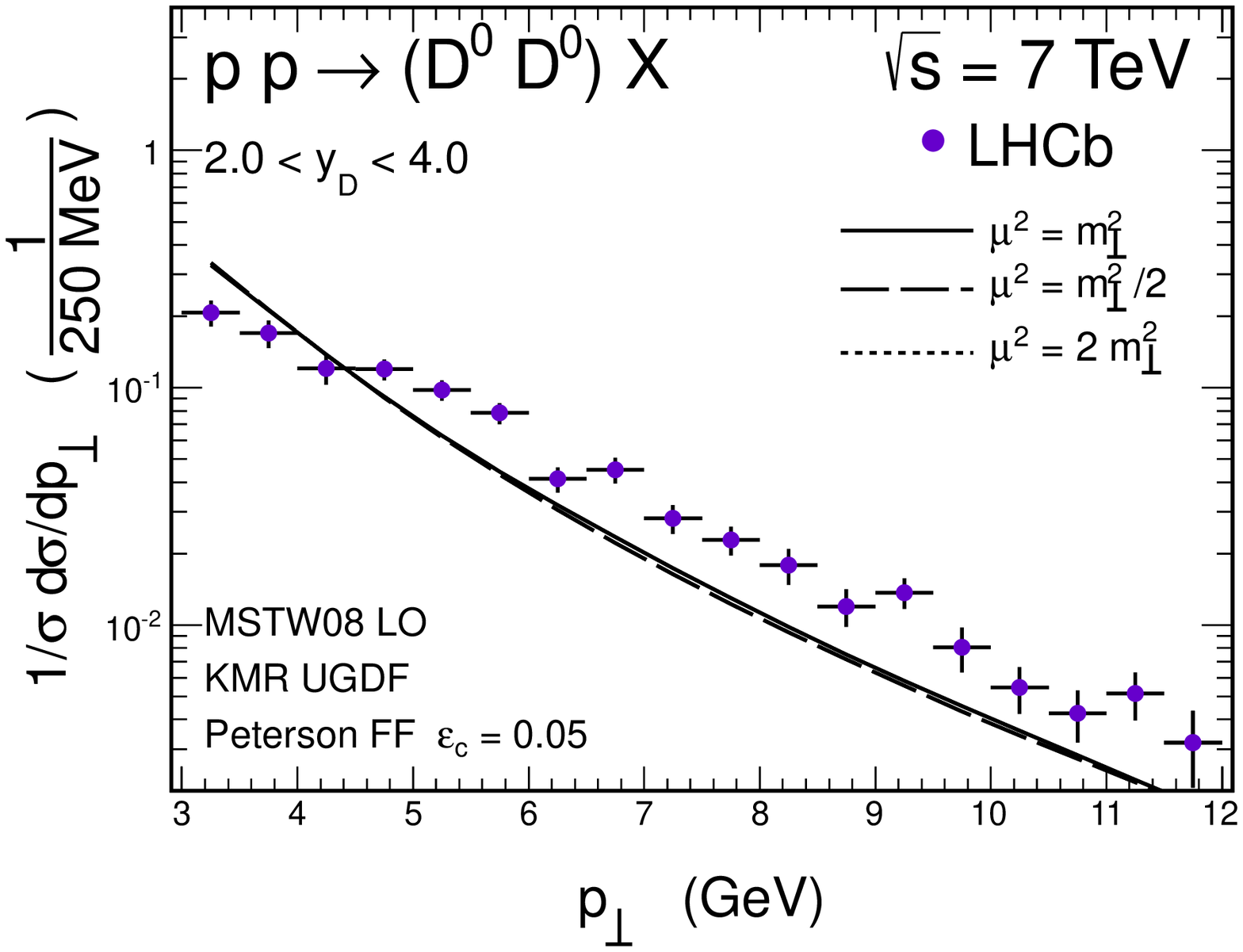}}
\end{minipage}
   \caption{
\small Transverse momentum distribution of $D^0$ mesons from the
$D^0 D^0$ pair contained in the LHCb kinematical region. The left panel shows
dependence on UGDFs, while the right panel illustrates dependence of the
result for the KMR UGDF on the factorization/renormalization scales.
}
\label{fig:pt-lhcb-DD-1}
\end{figure}

In Fig.~\ref{fig:pt-lhcb-DD-1} we present distribution in transverse
momentum of one of the $D^0$ mesons, provided that both are measured
within the LHCb experiment coverage specified in the figure caption.
Our theoretical distributions have shapes in rough agreement with the
experimental data. The shapes of the distributions are almost identical for 
different UGDFs used in the calculations (left panel) and are independent on the choice of scales
in the case of the KMR model (right panel).

%-----------------------------------------------------------------------------
%------------------------------------------------------------------------------

\begin{figure}[!h]
\begin{minipage}{0.47\textwidth}
 \centerline{\includegraphics[width=1.0\textwidth]{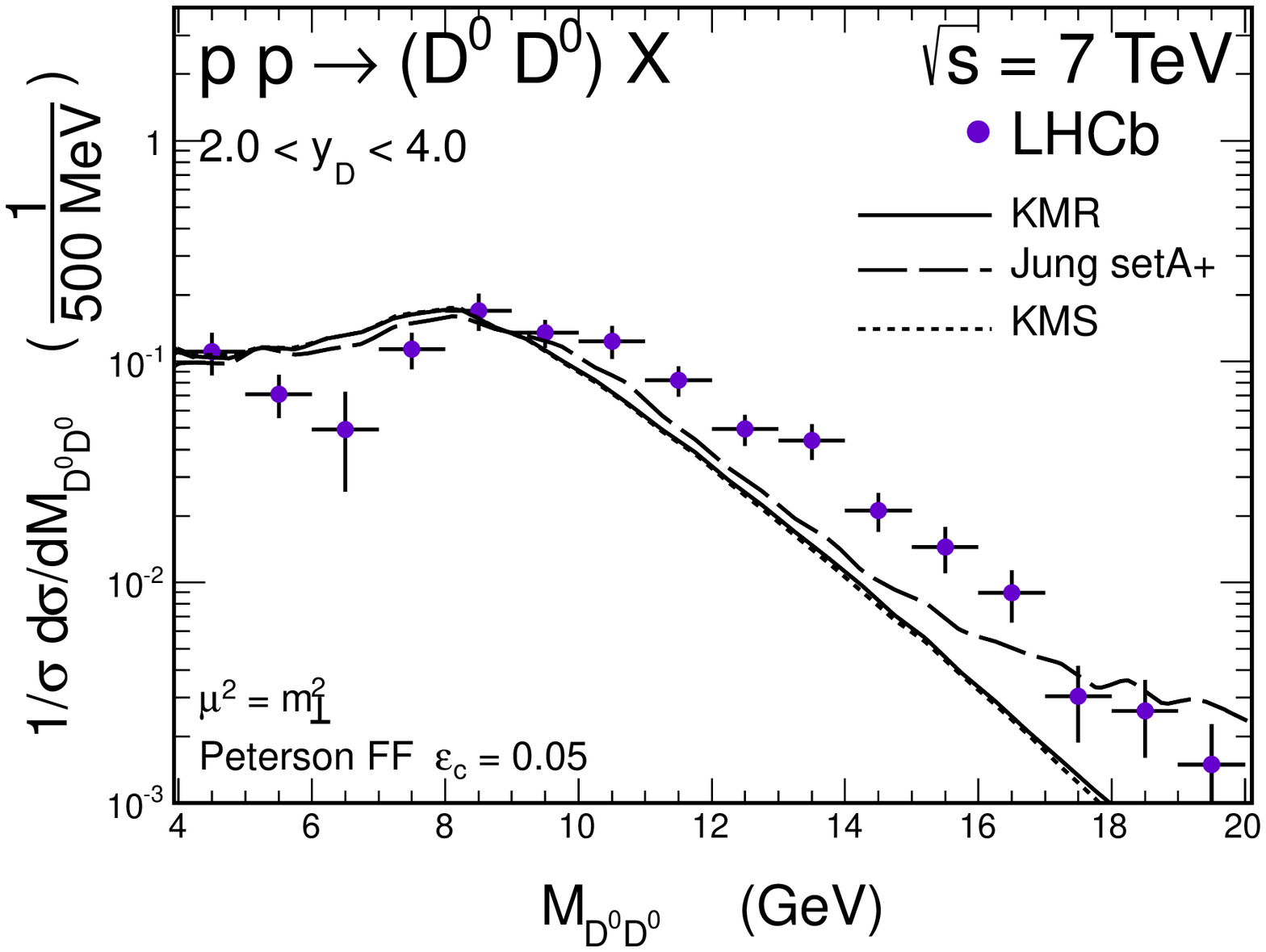}}
\end{minipage}
\hspace{0.5cm}
\begin{minipage}{0.47\textwidth}
 \centerline{\includegraphics[width=1.0\textwidth]{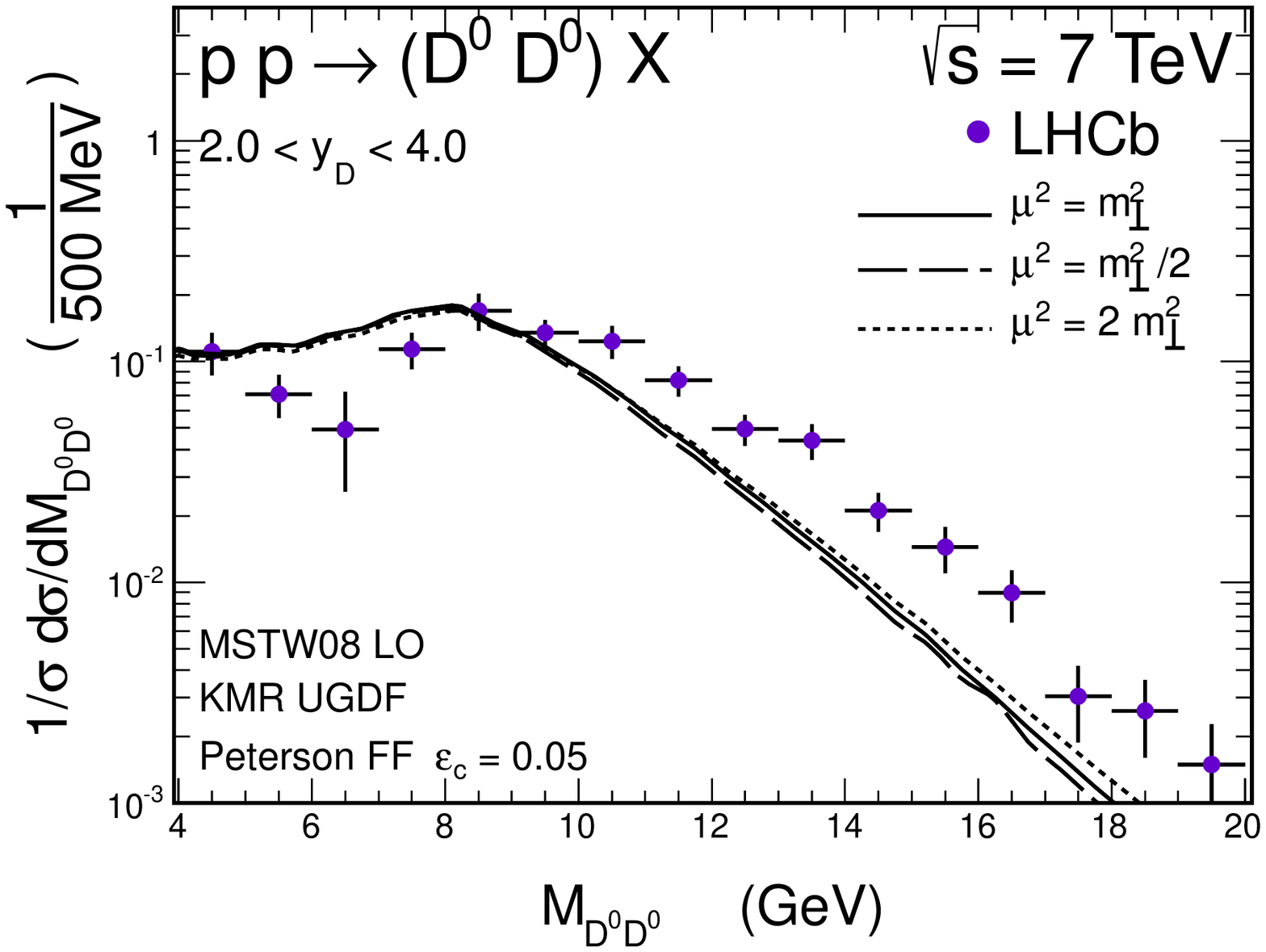}}
\end{minipage}
   \caption{
\small $M_{D^{0}D^{0}}$ invariant mass distribution for $D^0 D^0$ contained in the
LHCb kinematical region. The left panel shows
dependence on UGDFs, while the right panel illustrates dependence of the
result for the KMR UGDF on the factorization/renormalization scales.
}
\label{fig:minv-lhcb-DD-2}
\end{figure}

In Fig.~\ref{fig:minv-lhcb-DD-2} we show distribution in the $D^0 D^0$
invariant mass $M_{D^{0}D^{0}}$ for both $D^0$'s measured in the kinematical region
covered by the LHCb experiment. Here the shapes of the distributions have the same behavior
for various UGDFs and are insensitive to changes of scales as in the previous figure.
The characteristic minimum at small invariant masses is a consequnce of
experimental cuts (see Ref.~\cite{MS2013-charmed-meson}) and is rather well reproduced.

%-----------------------------------------------------------------------------
%------------------------------------------------------------------------------

\begin{figure}[!h]
\begin{minipage}{0.47\textwidth}
 \centerline{\includegraphics[width=1.0\textwidth]{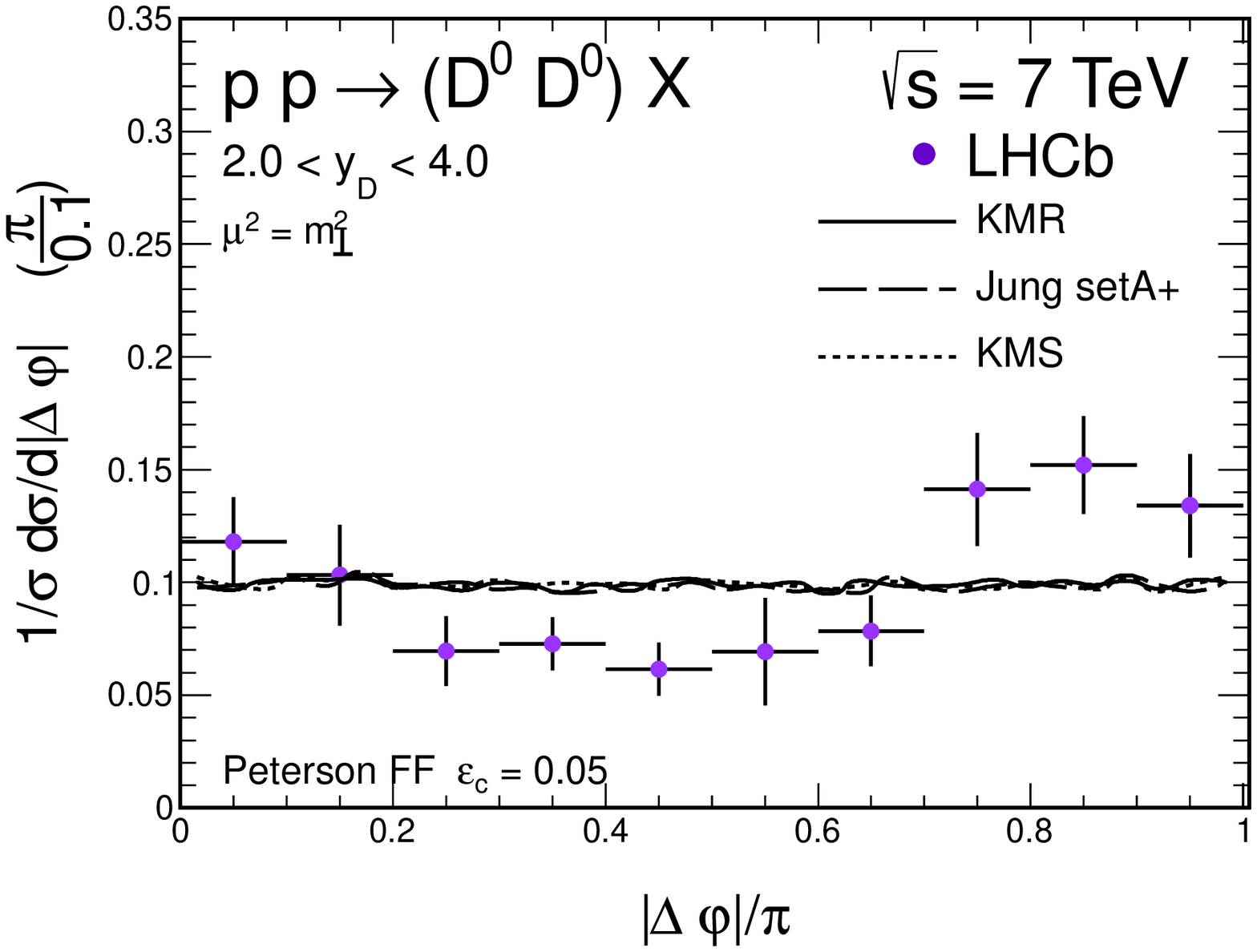}}
\end{minipage}
\hspace{0.5cm}
\begin{minipage}{0.47\textwidth}
 \centerline{\includegraphics[width=1.0\textwidth]{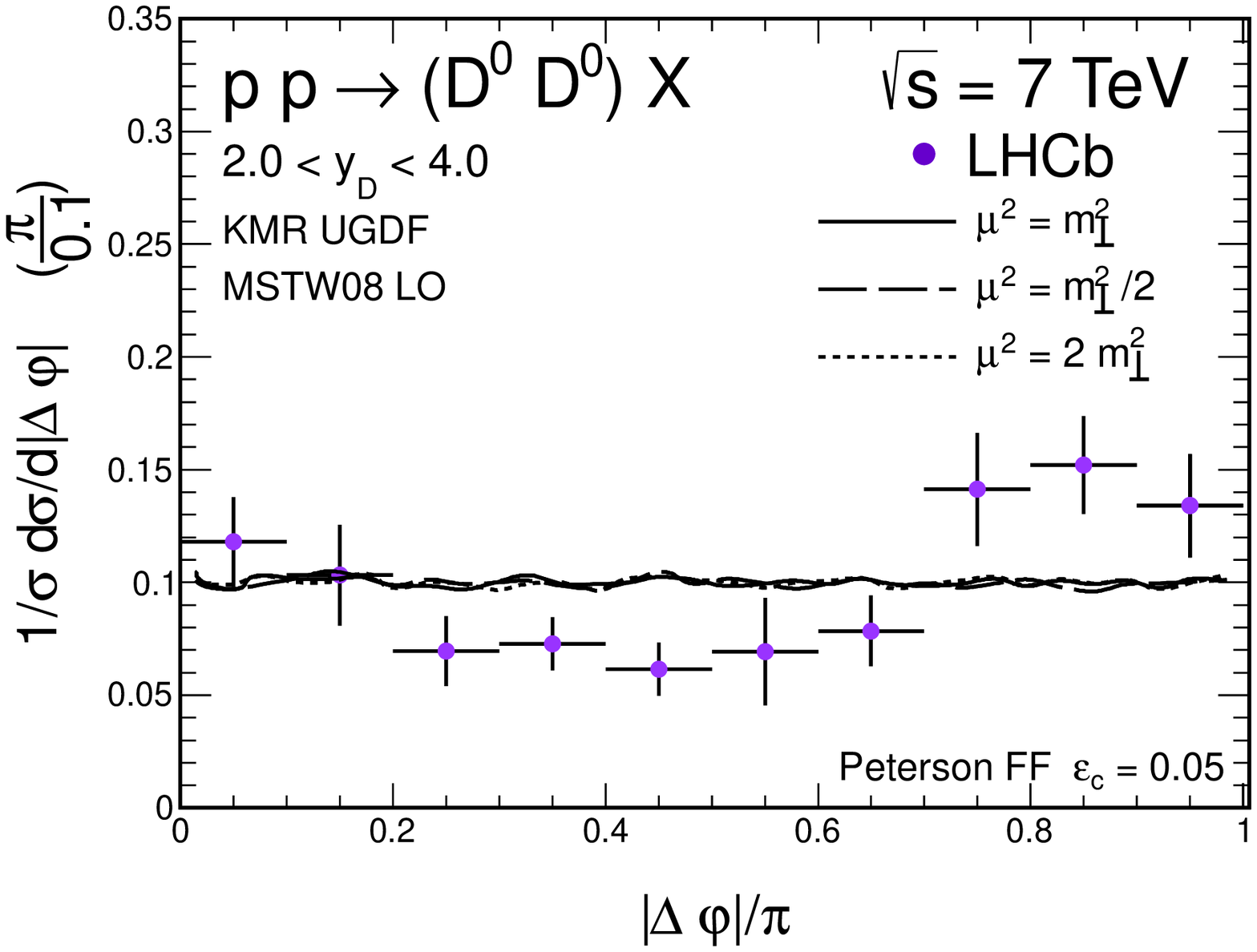}}
\end{minipage}
   \caption{
\small Distribution in azimuthal angle $\varphi_{D^{0}D^{0}}$ between both $D^0$'s.
The left panel shows
dependence on UGDFs, while the right panel illustrates dependence of the
result for the KMR UGDF on the factorization/renormalization scales.
}
\label{fig:phid-lhcb-DD-3}
\end{figure}

Finally in Fig.~\ref{fig:phid-lhcb-DD-3} we show distribution in
azimuthal angle $\varphi_{D^{0}D^{0}}$ between both $D^0$'s. While the theoretical DPS contribution is
independent of the relative azimuthal angle, there is some small
residual dependence on azimuthal angle in experimental distribution. 
This may show that there is some missing mechanism which gives contributions both at small and large $\Delta \varphi$. However, this discrepancy may be also an inherent property of the DPS factorized model which does not allow for any azimuthal correlations between particles produced in different hard scatterings.
We wish to emphasize in this context that the angular azimuthal correlation pattern for $D^{0}\bar{D^{0}}$, discussed in Ref.~\cite{MS2013-charmed-meson},
and for $D^{0}D^{0}$ $(\bar{D^{0}}\bar{D^{0}})$, discussed here, are quite different. The distribution for $D^{0}D^{0}$ $(\bar{D^{0}}\bar{D^{0}})$
is much more flat compared to the $D^{0}\bar{D^{0}}$ one which shows a pronounced maximum at $\varphi_{D^{0}\bar{D^{0}}} = 180^{\circ}$ (mostly from pair creation) and $\varphi_{D^{0}\bar{D^{0}}} = 0^{\circ}$ (mostly from gluon splitting) \cite{MS2013-charmed-meson}. This qualitative difference is in our opinion a model independent proof of the dominance of DPS effects in the production of $D^{0}D^{0}$ $(\bar{D^{0}}\bar{D^{0}})$.

%-----------------------------------------------------------------------------
%------------------------------------------------------------------------------

%------------------------------------------------------------
\subsection{DPS \bm{$c \bar c c \bar c$} production and inclusive
charmed meson distributions}
%------------------------------------------------------------

Since the DPS cross section is very large
it is also very important to look at the DPS $c \bar c c \bar c$
contribution to inclusive charmed meson spectra.
Let us consider for example transverse momentum distribution
of a charmed $D_i$ meson. The corresponding DPS $c \bar c c \bar c$ 
contribution can be written as:

\begin{eqnarray}
\frac{d\sigma_{inc}^{D_{i}, DPS}}{dp_t} =  \nonumber
P_{D_i} (1 - P_{D_i}) \frac{d \sigma^D}{d p_{1,t}}|_{p_{1,t}=p_{t}}(-2.1 < \eta_1 < 2.1, -\infty < \eta_2 < \infty)\\ \nonumber + \;\;
P_{D_i} (1 - P_{D_i}) \frac{d \sigma^D}{d p_{2,t}}|_{p_{2,t}=p_{t}}(-\infty < \eta_1 < \infty, -2.1 < \eta_2 < 2.1)\\ \nonumber + \;\; 
P_{D_i} P_{D_i}  \frac{d \sigma^D}{d p_{1,t}}|_{p_{1,t}=p_{t}}(-2.1 < \eta_1 < 2.1, -\infty < \eta_2 < \infty)\\  + \;\; 
P_{D_i} P_{D_i} \frac{d \sigma^D}{d p_{2,t}}|_{p_{2,t}=p_{t}}(-\infty < \eta_1 < \infty, -2.1 < \eta_2 < 2.1). 
\label{inclusive_dsig_dpt_D}
\end{eqnarray}

In the formula above $P_{D_i}$ is a shorthand notation for the
branching fraction $P_{c \to D_i}$ and $\sigma^{D}$ is the cross section
for $D$-mesons assuming artificially the branching fraction equal to 1.
The formula above can be somewhat simplified when combining similar terms.

%-----------------------------------------------------------------------------
\begin{figure}[!h]
\begin{minipage}{0.47\textwidth}
 \centerline{\includegraphics[width=1.0\textwidth]{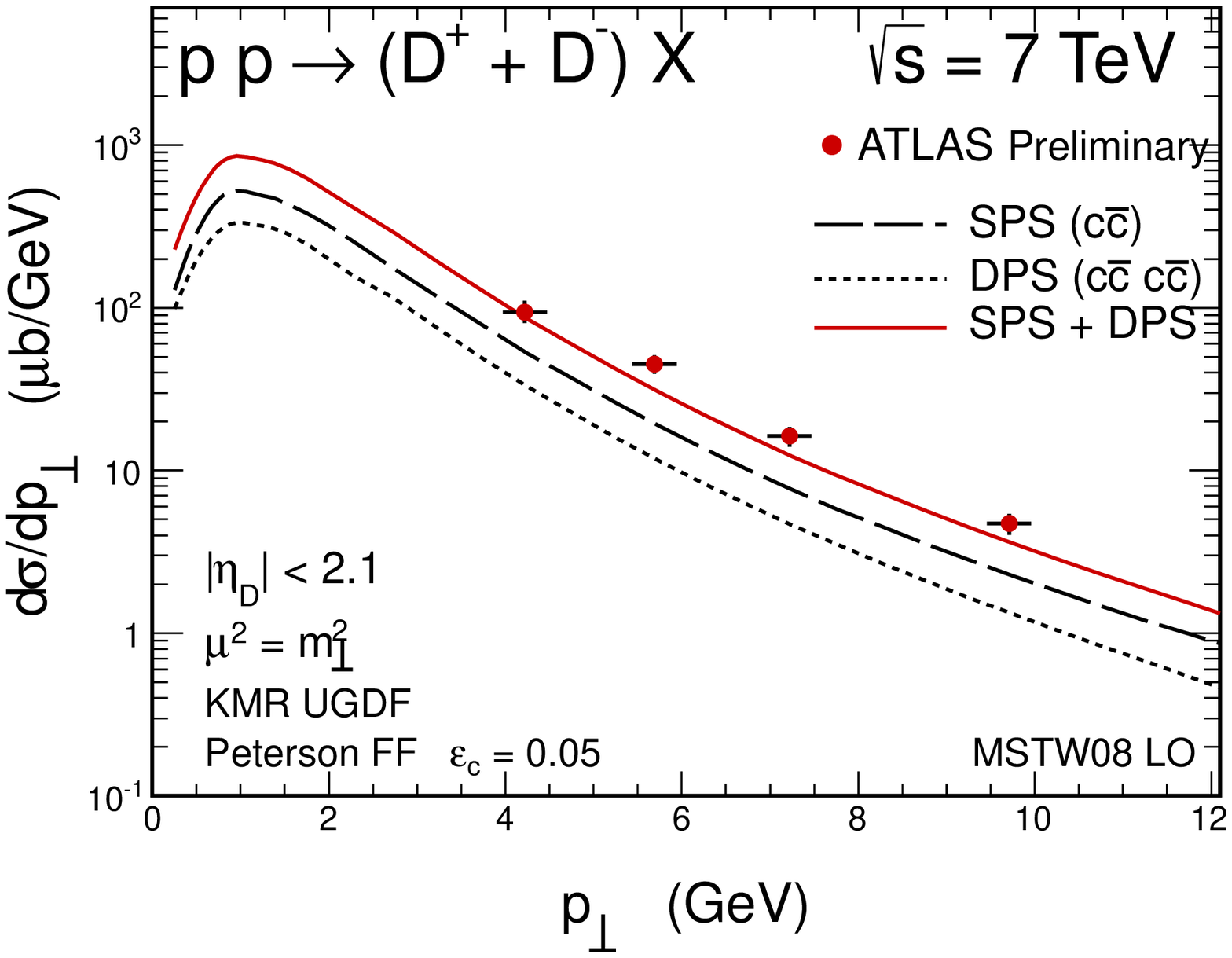}}
\end{minipage}
\hspace{0.5cm}
\begin{minipage}{0.47\textwidth}
 \centerline{\includegraphics[width=1.0\textwidth]{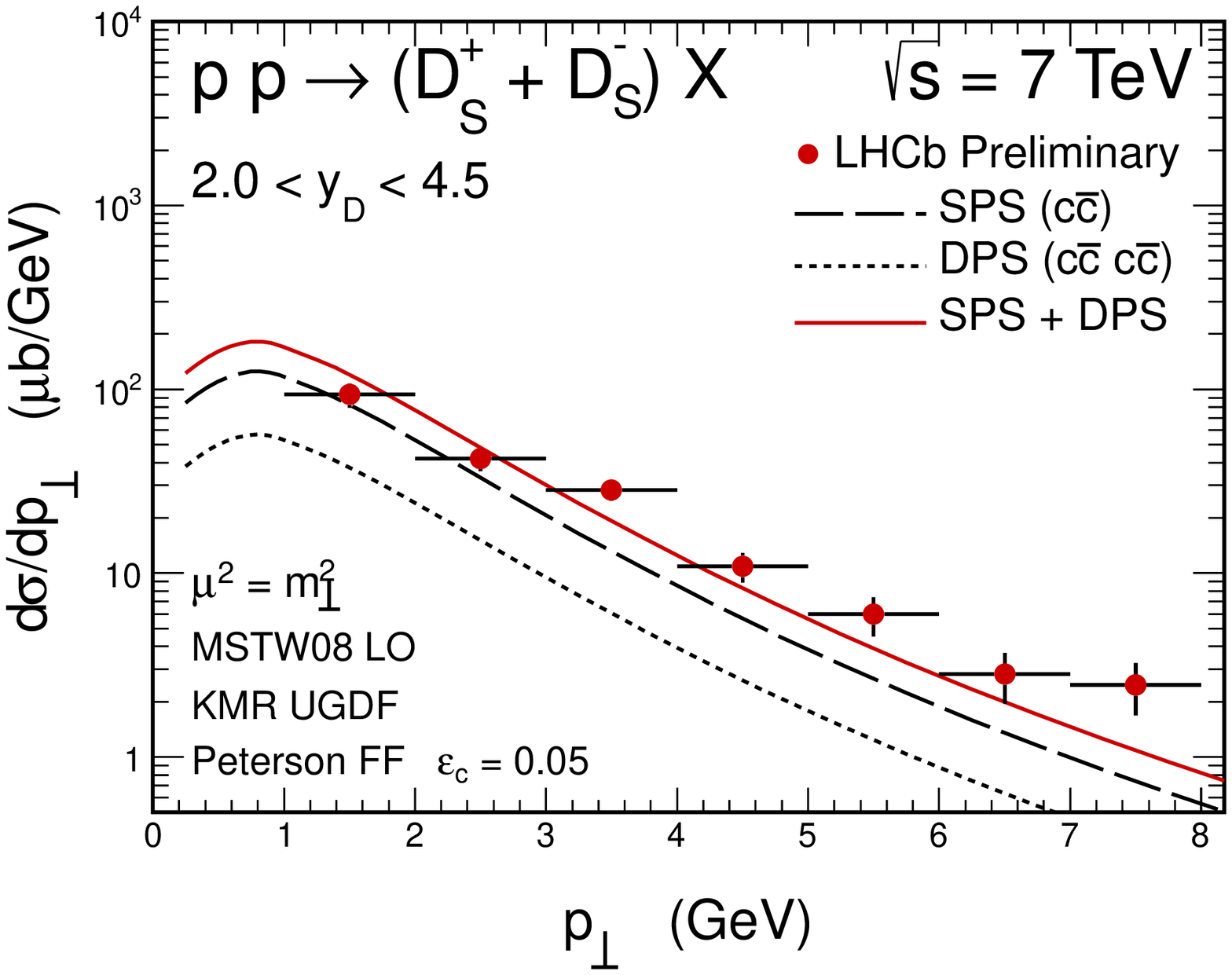}}
\end{minipage}
\begin{minipage}{0.47\textwidth}
 \centerline{\includegraphics[width=1.0\textwidth]{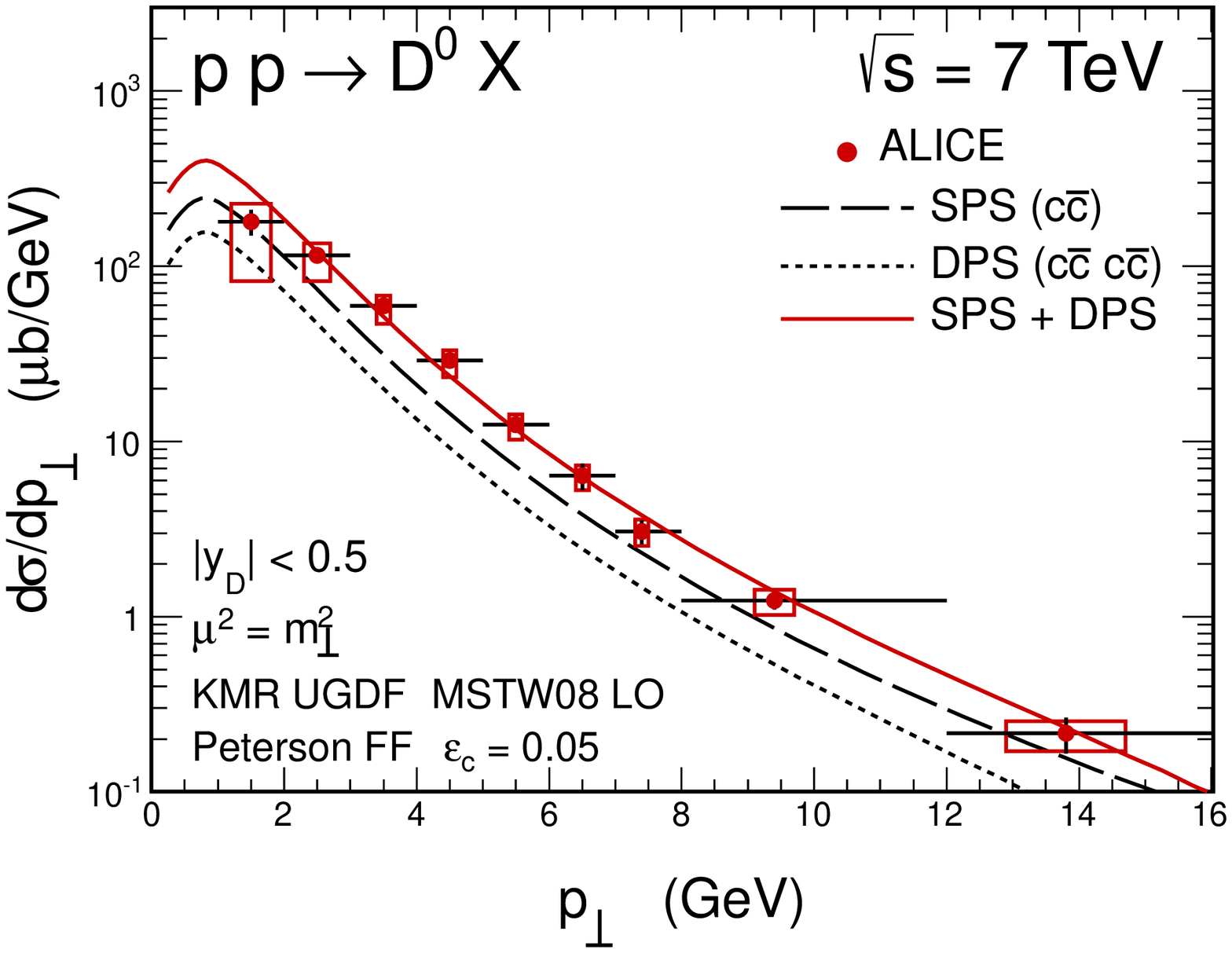}}
\end{minipage}
   \caption{
\small Inclusive transverse momentum distributions of different charmed mesons
measured by different groups at the LHC. The long-dashed line corresponds to
 the standard SPS $c \bar c$ production and the dotted line represents the DPS
 $c \bar c c \bar c$ contribution.
}
\label{fig:pt-inc-atlas-lhcb-alice}
\end{figure}
%------------------------------------------------------------------------------

In Fig.~\ref{fig:pt-inc-atlas-lhcb-alice} we show inclusive one pair (long-dashed line),
inclusive DPS two-pair contribution (dotted line) and the sum of both terms
to transverse momentum distribution of different $D$ mesons (solid line). The DPS 
$c \bar c c \bar c$ contribution is of the same order as the standard
traditional SPS $c \bar c$ contribution. This is a completely new situation
compared to what it was at smaller energies. The sum of both
contributions almost describes the different experimental data.
As discussed in the previous section the SPS $c \bar c c \bar c$
contribution can be of the order of $10\%$ of the DPS $c \bar c c \bar c$ contribution.
At higher energies one could expect even relatively larger DPS 
$c \bar c c \bar c$  contribution. A problem could start, however, that then
one enters the region of really small gluon longitudinal momentum fractions 
$x <$ 10$^{-4}$ for which the gluon UGDFs (or PDFs) are not well known. 
In this case realistic models of UGDFs are badly needed. Do we have such a
distribution at present? 

%--------------------------
\section{Conclusions}
%--------------------------

In this paper we have discussed production of $c \bar c c \bar c$
in the double-parton scattering (DPS) and single-parton scattering (SPS)
in the $g g \to c \bar c c \bar c$ subprocess. The double-parton
scattering is calculated in the factorized Ansatz with each step
calculated in the $k_t$-factorization approach, i.e. including
effectively higher-order QCD corrections.

The cross section in the $k_t$-factorization approach turned out
to be much larger than its counterpart calculated in the LO
collinear approach. The distribution in rapidity difference between
quarks/antiquarks from the same and different scatterings turned
out to have similar shape as in the LO collinear approach. The same
is true for invariant masses of pairs of quark-quark, antiquark-antiquark and quark-antiquark, etc.
The distribution in transverse momentum of the pair from the same
scattering turned out to be similar to that for the pairs originating from
different scatterings. 

We have calculated also cross sections for the production of 
$D_i D_j$ (both containing $c$ quark)
and $\bar D_i \bar D_j$ (both containing $\bar c$ antiquark) pairs of mesons.
The results of the calculation have been compared to recent results of
the LHCb collaboration.

The total rates of the meson pair production depend on the unintegrated
gluon distributions. The best agreement with the LHCb result has been 
obtained for the Kimber-Martin-Ryskin UGDF. This approach as discussed already in the literature 
effectively includes higher-order QCD corrections.

As an example we have also calculated several differential distributions
for $D^0 D^0$ pair production. Rather good agreement has been obtained
for transverse momentum distribution of $D^0$ $(\bar D^0)$ mesons
and $D^0 D^0$ invariant mass distribution. The distribution in azimuthal
angle between both $D^0$'s suggests that some contributions may be still
missing. The single parton scattering contribution, calculated in the
high energy approximation, turned out to be rather small. This should
be checked in exact $2 \to 4$ parton model calculations in the future. 

We have shown that the DPS mechanism of $c \bar c c
\bar c$ production gives a new significant contribution to inclusive
charmed meson spectra. For instance the description of the inclusive ATLAS, ALICE and LHCb data is
very difficult in terms of the conventional SPS ($c \bar c$) contribution \cite{MS2013-charmed-meson}.

Since we have shown that the DPS mechanism gives significant
contribution to inclusive spectra of charmed mesons the estimate of 
DPS effects, presented in Ref.~\cite{Berezhnoy2012} and based on experimental 
inclusive cross section, leads to an overestimation of the DPS effect.

Summarizing, the present study of $c \bar c c \bar c$ reaction in the
$k_t$-factorization approach has shown that this reaction is an extremely good testing
ground of double-parton scattering effects. 
The LHCb kinematics is not the best in this respect.
Both ATLAS and CMS collaborations could measure the production of pairs 
of $D_i D_j$ and/or $\bar D_i \bar D_j$ mesons with large rapidity distance 
where the DPS mechanism is predicted to clearly dominate over the SPS 
mechanism. Another potentially interesting place to investigate DPS
effect is the $p p \to J/\psi J/\psi X$ reaction \cite{BSZSS2012}.
Similarly as for $p p \to c \bar c c \bar c X$ discussed here, the large
rapidity gap between two $J/\psi$'s should select clear sample of DPS 
mechanism.

\vspace{1cm}

{\bf Acknowledgments}

The authors thank Ivan Belyaev and Marek Szczekowski for useful discussions of many 
aspects of the LHCb experimental data and Wolfgang Sch{\"a}fer for discussion 
of single parton scattering contribution.
This work is supported in part by the Polish
Grants DEC-2011/01/B/ST2/04535 and N202 237040.

%-------------------------------------------------------------------------------------

\end{document}